\documentclass[twocolumn,traditabstract,longauth]{aa}

\def\setsymbol#1#2{\expandafter\def\csname #1\endcsname{#2}}
\def\getsymbol#1{\csname #1\endcsname}

\def\Planck{\textit{Planck}}

\def\all2013resultspapers{\nocite{planck2013-p01, planck2013-p02, planck2013-p02a, planck2013-p02d, planck2013-p02b, planck2013-p03, planck2013-p03c, planck2013-p03f, planck2013-p03d, planck2013-p03e, planck2013-p01a, planck2013-p06, planck2013-p03a, planck2013-pip88, planck2013-p08, planck2013-p11, planck2013-p12, planck2013-p13, planck2013-p14, planck2013-p15, planck2013-p05b, planck2013-p17, planck2013-p09, planck2013-p09a, planck2013-p20, planck2013-p19, planck2013-pipaberration, planck2013-p05, planck2013-p05a, planck2013-pip56, planck2013-p06b}}

\newbox\tablebox    \newdimen\tablewidth
\def\leaderfil{\leaders\hbox to 5pt{\hss.\hss}\hfil}
\def\endPlancktable{\tablewidth=\columnwidth 
    $$\hss\copy\tablebox\hss$$
    \vskip-\lastskip\vskip -2pt}

\def\tablenote#1 #2\par{\begingroup \parindent=0.8em
    \abovedisplayshortskip=0pt\belowdisplayshortskip=0pt
    \noindent
    $$\hss\vbox{\hsize\tablewidth \hangindent=\parindent \hangafter=1 \noindent
    \hbox to \parindent{$^#1$\hss}\strut#2\strut\par}\hss$$
    \endgroup}
\def\doubleline{\vskip 3pt\hrule \vskip 1.5pt \hrule \vskip 5pt}

\def\L2{\ifmmode L_2\else $L_2$\fi}

\def\DeltaT{\ifmmode \Delta T\else $\Delta T$\fi}
\def\deltat{\ifmmode \Delta t\else $\Delta t$\fi}
\def\fknee{\ifmmode f_{\rm knee}\else $f_{\rm knee}$\fi}
\def\Fmax{\ifmmode F_{\rm max}\else $F_{\rm max}$\fi}
\def\solar{\ifmmode{\rm M}_{\mathord\odot}\else${\rm M}_{\mathord\odot}$\fi}
\def\Msolar{\ifmmode{\rm M}_{\mathord\odot}\else${\rm M}_{\mathord\odot}$\fi}
\def\Lsolar{\ifmmode{\rm L}_{\mathord\odot}\else${\rm L}_{\mathord\odot}$\fi}

\def\inv{\ifmmode^{-1}\else$^{-1}$\fi}
\def\mo{\ifmmode^{-1}\else$^{-1}$\fi}
\def\sup#1{\ifmmode ^{\rm #1}\else $^{\rm #1}$\fi}
\def\expo#1{\ifmmode \times 10^{#1}\else $\times 10^{#1}$\fi}
\def\,{\thinspace}
\def\lsim{\mathrel{\raise .4ex\hbox{\rlap{$<$}\lower 1.2ex\hbox{$\sim$}}}}
\def\gsim{\mathrel{\raise .4ex\hbox{\rlap{$>$}\lower 1.2ex\hbox{$\sim$}}}}

\def\simprop{\mathrel{\raise .4ex\hbox{\rlap{$\propto$}\lower 1.2ex\hbox{$\sim$}}}}
\def\deg{\ifmmode^\circ\else$^\circ$\fi}
\def\pdeg{\ifmmode $\setbox0=\hbox{$^{\circ}$}\rlap{\hskip.11\wd0 .}$^{\circ}
          \else \setbox0=\hbox{$^{\circ}$}\rlap{\hskip.11\wd0 .}$^{\circ}$\fi}
\def\arcs{\ifmmode {^{\scriptstyle\prime\prime}}
          \else $^{\scriptstyle\prime\prime}$\fi}
\def\arcm{\ifmmode {^{\scriptstyle\prime}}
          \else $^{\scriptstyle\prime}$\fi}
\newdimen\sa  \newdimen\sb
\def\parcs{\sa=.07em \sb=.03em
     \ifmmode \hbox{\rlap{.}}^{\scriptstyle\prime\kern -\sb\prime}\hbox{\kern -\sa}
     \else \rlap{.}$^{\scriptstyle\prime\kern -\sb\prime}$\kern -\sa\fi}
\def\parcm{\sa=.08em \sb=.03em
     \ifmmode \hbox{\rlap{.}\kern\sa}^{\scriptstyle\prime}\hbox{\kern-\sb}
     \else \rlap{.}\kern\sa$^{\scriptstyle\prime}$\kern-\sb\fi}
\def\ra[#1 #2 #3.#4]{#1\sup{h}#2\sup{m}#3\sup{s}\llap.#4}
\def\dec[#1 #2 #3.#4]{#1\deg#2\arcm#3\arcs\llap.#4}
\def\deco[#1 #2 #3]{#1\deg#2\arcm#3\arcs}
\def\rra[#1 #2]{#1\sup{h}#2\sup{m}}

\def\dots{\relax\ifmmode \ldots\else $\ldots$\fi}
\def\WHzsr{\ifmmode $W\,Hz\mo\,sr\mo$\else W\,Hz\mo\,sr\mo\fi}
\def\mHz{\ifmmode $\,mHz$\else \,mHz\fi}
\def\GHz{\ifmmode $\,GHz$\else \,GHz\fi}
\def\mKs{\ifmmode $\,mK\,s$^{1/2}\else \,mK\,s$^{1/2}$\fi}
\def\muKs{\ifmmode \,\mu$K\,s$^{1/2}\else \,$\mu$K\,s$^{1/2}$\fi}
\def\muKRJs{\ifmmode \,\mu$K$_{\rm RJ}$\,s$^{1/2}\else \,$\mu$K$_{\rm RJ}$\,s$^{1/2}$\fi}
\def\muKHz{\ifmmode \,\mu$K\,Hz$^{-1/2}\else \,$\mu$K\,Hz$^{-1/2}$\fi}
\def\MJysr{\ifmmode \,$MJy\,sr\mo$\else \,MJy\,sr\mo\fi}
\def\MJysrmK{\ifmmode \,$MJy\,sr\mo$\,mK$_{\rm CMB}\mo\else \,MJy\,sr\mo\,mK$_{\rm CMB}\mo$\fi}
\def\microns{\ifmmode \,\mu$m$\else \,$\mu$m\fi}
\def\micron{\microns}
\def\muK{\ifmmode \,\mu$K$\else \,$\mu$\hbox{K}\fi}
\def\microK{\ifmmode \,\mu$K$\else \,$\mu$\hbox{K}\fi}
\def\muW{\ifmmode \,\mu$W$\else \,$\mu$\hbox{W}\fi}
\def\kms{\ifmmode $\,km\,s$^{-1}\else \,km\,s$^{-1}$\fi}
\def\kmsMpc{\ifmmode $\,\kms\,Mpc\mo$\else \,\kms\,Mpc\mo\fi}
\setsymbol{LFI:center:frequency:70GHz:units}{70.3\,GHz}
\setsymbol{LFI:center:frequency:44GHz:units}{44.1\,GHz}
\setsymbol{LFI:center:frequency:30GHz:units}{28.5\,GHz}

\setsymbol{LFI:center:frequency:70GHz}{70.3}
\setsymbol{LFI:center:frequency:44GHz}{44.1}
\setsymbol{LFI:center:frequency:30GHz}{28.5}

\setsymbol{LFI:center:frequency:LFI18:Rad:M:units}{71.7\GHz}
\setsymbol{LFI:center:frequency:LFI19:Rad:M:units}{67.5\GHz}
\setsymbol{LFI:center:frequency:LFI20:Rad:M:units}{69.2\GHz}
\setsymbol{LFI:center:frequency:LFI21:Rad:M:units}{70.4\GHz}
\setsymbol{LFI:center:frequency:LFI22:Rad:M:units}{71.5\GHz}
\setsymbol{LFI:center:frequency:LFI23:Rad:M:units}{70.8\GHz}
\setsymbol{LFI:center:frequency:LFI24:Rad:M:units}{44.4\GHz}
\setsymbol{LFI:center:frequency:LFI25:Rad:M:units}{44.0\GHz}
\setsymbol{LFI:center:frequency:LFI26:Rad:M:units}{43.9\GHz}
\setsymbol{LFI:center:frequency:LFI27:Rad:M:units}{28.3\GHz}
\setsymbol{LFI:center:frequency:LFI28:Rad:M:units}{28.8\GHz}
\setsymbol{LFI:center:frequency:LFI18:Rad:S:units}{70.1\GHz}
\setsymbol{LFI:center:frequency:LFI19:Rad:S:units}{69.6\GHz}
\setsymbol{LFI:center:frequency:LFI20:Rad:S:units}{69.5\GHz}
\setsymbol{LFI:center:frequency:LFI21:Rad:S:units}{69.5\GHz}
\setsymbol{LFI:center:frequency:LFI22:Rad:S:units}{72.8\GHz}
\setsymbol{LFI:center:frequency:LFI23:Rad:S:units}{71.3\GHz}
\setsymbol{LFI:center:frequency:LFI24:Rad:S:units}{44.1\GHz}
\setsymbol{LFI:center:frequency:LFI25:Rad:S:units}{44.1\GHz}
\setsymbol{LFI:center:frequency:LFI26:Rad:S:units}{44.1\GHz}
\setsymbol{LFI:center:frequency:LFI27:Rad:S:units}{28.5\GHz}
\setsymbol{LFI:center:frequency:LFI28:Rad:S:units}{28.2\GHz}

\setsymbol{LFI:center:frequency:LFI18:Rad:M}{71.7}
\setsymbol{LFI:center:frequency:LFI19:Rad:M}{67.5}
\setsymbol{LFI:center:frequency:LFI20:Rad:M}{69.2}
\setsymbol{LFI:center:frequency:LFI21:Rad:M}{70.4}
\setsymbol{LFI:center:frequency:LFI22:Rad:M}{71.5}
\setsymbol{LFI:center:frequency:LFI23:Rad:M}{70.8}
\setsymbol{LFI:center:frequency:LFI24:Rad:M}{44.4}
\setsymbol{LFI:center:frequency:LFI25:Rad:M}{44.0}
\setsymbol{LFI:center:frequency:LFI26:Rad:M}{43.9}
\setsymbol{LFI:center:frequency:LFI27:Rad:M}{28.3}
\setsymbol{LFI:center:frequency:LFI28:Rad:M}{28.8}
\setsymbol{LFI:center:frequency:LFI18:Rad:S}{70.1}
\setsymbol{LFI:center:frequency:LFI19:Rad:S}{69.6}
\setsymbol{LFI:center:frequency:LFI20:Rad:S}{69.5}
\setsymbol{LFI:center:frequency:LFI21:Rad:S}{69.5}
\setsymbol{LFI:center:frequency:LFI22:Rad:S}{72.8}
\setsymbol{LFI:center:frequency:LFI23:Rad:S}{71.3}
\setsymbol{LFI:center:frequency:LFI24:Rad:S}{44.1}
\setsymbol{LFI:center:frequency:LFI25:Rad:S}{44.1}
\setsymbol{LFI:center:frequency:LFI26:Rad:S}{44.1}
\setsymbol{LFI:center:frequency:LFI27:Rad:S}{28.5}
\setsymbol{LFI:center:frequency:LFI28:Rad:S}{28.2}

\setsymbol{LFI:white:noise:sensitivity:70GHz:units}{134.7\muKs}
\setsymbol{LFI:white:noise:sensitivity:44GHz:units}{164.7\muKs}
\setsymbol{LFI:white:noise:sensitivity:30GHz:units}{143.4\muKs}

\setsymbol{LFI:white:noise:sensitivity:70GHz}{134.7}
\setsymbol{LFI:white:noise:sensitivity:44GHz}{164.7}
\setsymbol{LFI:white:noise:sensitivity:30GHz}{143.4}

\setsymbol{LFI:white:noise:sensitivity:LFI18:Rad:M:units}{512.0\muKs}
\setsymbol{LFI:white:noise:sensitivity:LFI19:Rad:M:units}{581.4\muKs}
\setsymbol{LFI:white:noise:sensitivity:LFI20:Rad:M:units}{590.8\muKs}
\setsymbol{LFI:white:noise:sensitivity:LFI21:Rad:M:units}{455.2\muKs}
\setsymbol{LFI:white:noise:sensitivity:LFI22:Rad:M:units}{492.0\muKs}
\setsymbol{LFI:white:noise:sensitivity:LFI23:Rad:M:units}{507.7\muKs}
\setsymbol{LFI:white:noise:sensitivity:LFI24:Rad:M:units}{462.2\muKs}
\setsymbol{LFI:white:noise:sensitivity:LFI25:Rad:M:units}{413.6\muKs}
\setsymbol{LFI:white:noise:sensitivity:LFI26:Rad:M:units}{478.6\muKs}
\setsymbol{LFI:white:noise:sensitivity:LFI27:Rad:M:units}{277.7\muKs}
\setsymbol{LFI:white:noise:sensitivity:LFI28:Rad:M:units}{312.3\muKs}
\setsymbol{LFI:white:noise:sensitivity:LFI18:Rad:S:units}{465.7\muKs}
\setsymbol{LFI:white:noise:sensitivity:LFI19:Rad:S:units}{555.6\muKs}
\setsymbol{LFI:white:noise:sensitivity:LFI20:Rad:S:units}{623.2\muKs}
\setsymbol{LFI:white:noise:sensitivity:LFI21:Rad:S:units}{564.1\muKs}
\setsymbol{LFI:white:noise:sensitivity:LFI22:Rad:S:units}{534.4\muKs}
\setsymbol{LFI:white:noise:sensitivity:LFI23:Rad:S:units}{542.4\muKs}
\setsymbol{LFI:white:noise:sensitivity:LFI24:Rad:S:units}{399.2\muKs}
\setsymbol{LFI:white:noise:sensitivity:LFI25:Rad:S:units}{392.6\muKs}
\setsymbol{LFI:white:noise:sensitivity:LFI26:Rad:S:units}{418.6\muKs}
\setsymbol{LFI:white:noise:sensitivity:LFI27:Rad:S:units}{302.9\muKs}
\setsymbol{LFI:white:noise:sensitivity:LFI28:Rad:S:units}{285.3\muKs}

\setsymbol{LFI:white:noise:sensitivity:LFI18:Rad:M}{512.0}
\setsymbol{LFI:white:noise:sensitivity:LFI19:Rad:M}{581.4}
\setsymbol{LFI:white:noise:sensitivity:LFI20:Rad:M}{590.8}
\setsymbol{LFI:white:noise:sensitivity:LFI21:Rad:M}{455.2}
\setsymbol{LFI:white:noise:sensitivity:LFI22:Rad:M}{492.0}
\setsymbol{LFI:white:noise:sensitivity:LFI23:Rad:M}{507.7}
\setsymbol{LFI:white:noise:sensitivity:LFI24:Rad:M}{462.2}
\setsymbol{LFI:white:noise:sensitivity:LFI25:Rad:M}{413.6}
\setsymbol{LFI:white:noise:sensitivity:LFI26:Rad:M}{478.6}
\setsymbol{LFI:white:noise:sensitivity:LFI27:Rad:M}{277.7}
\setsymbol{LFI:white:noise:sensitivity:LFI28:Rad:M}{312.3}
\setsymbol{LFI:white:noise:sensitivity:LFI18:Rad:S}{465.7}
\setsymbol{LFI:white:noise:sensitivity:LFI19:Rad:S}{555.6}
\setsymbol{LFI:white:noise:sensitivity:LFI20:Rad:S}{623.2}
\setsymbol{LFI:white:noise:sensitivity:LFI21:Rad:S}{564.1}
\setsymbol{LFI:white:noise:sensitivity:LFI22:Rad:S}{534.4}
\setsymbol{LFI:white:noise:sensitivity:LFI23:Rad:S}{542.4}
\setsymbol{LFI:white:noise:sensitivity:LFI24:Rad:S}{399.2}
\setsymbol{LFI:white:noise:sensitivity:LFI25:Rad:S}{392.6}
\setsymbol{LFI:white:noise:sensitivity:LFI26:Rad:S}{418.6}
\setsymbol{LFI:white:noise:sensitivity:LFI27:Rad:S}{302.9}
\setsymbol{LFI:white:noise:sensitivity:LFI28:Rad:S}{285.3}

\setsymbol{LFI:knee:frequency:70GHz:units}{29.5\mHz}
\setsymbol{LFI:knee:frequency:44GHz:units}{56.2\mHz}
\setsymbol{LFI:knee:frequency:30GHz:units}{113.7\mHz}

\setsymbol{LFI:knee:frequency:70GHz}{29.5}
\setsymbol{LFI:knee:frequency:44GHz}{56.2}
\setsymbol{LFI:knee:frequency:30GHz}{113.7}

\setsymbol{LFI:knee:frequency:LFI18:Rad:M:units}{16.3\mHz}
\setsymbol{LFI:knee:frequency:LFI19:Rad:M:units}{15.1\mHz}
\setsymbol{LFI:knee:frequency:LFI20:Rad:M:units}{18.7\mHz}
\setsymbol{LFI:knee:frequency:LFI21:Rad:M:units}{37.2\mHz}
\setsymbol{LFI:knee:frequency:LFI22:Rad:M:units}{12.7\mHz}
\setsymbol{LFI:knee:frequency:LFI23:Rad:M:units}{34.6\mHz}
\setsymbol{LFI:knee:frequency:LFI24:Rad:M:units}{46.2\mHz}
\setsymbol{LFI:knee:frequency:LFI25:Rad:M:units}{24.9\mHz}
\setsymbol{LFI:knee:frequency:LFI26:Rad:M:units}{67.6\mHz}
\setsymbol{LFI:knee:frequency:LFI27:Rad:M:units}{187.4\mHz}
\setsymbol{LFI:knee:frequency:LFI28:Rad:M:units}{122.2\mHz}
\setsymbol{LFI:knee:frequency:LFI18:Rad:S:units}{17.7\mHz}
\setsymbol{LFI:knee:frequency:LFI19:Rad:S:units}{22.0\mHz}
\setsymbol{LFI:knee:frequency:LFI20:Rad:S:units}{8.7\mHz}
\setsymbol{LFI:knee:frequency:LFI21:Rad:S:units}{25.9\mHz}
\setsymbol{LFI:knee:frequency:LFI22:Rad:S:units}{15.8\mHz}
\setsymbol{LFI:knee:frequency:LFI23:Rad:S:units}{129.8\mHz}
\setsymbol{LFI:knee:frequency:LFI24:Rad:S:units}{100.9\mHz}
\setsymbol{LFI:knee:frequency:LFI25:Rad:S:units}{38.9\mHz}
\setsymbol{LFI:knee:frequency:LFI26:Rad:S:units}{58.9\mHz}
\setsymbol{LFI:knee:frequency:LFI27:Rad:S:units}{104.4\mHz}
\setsymbol{LFI:knee:frequency:LFI28:Rad:S:units}{40.7\mHz}

\setsymbol{LFI:knee:frequency:LFI18:Rad:M}{16.3}
\setsymbol{LFI:knee:frequency:LFI19:Rad:M}{15.1}
\setsymbol{LFI:knee:frequency:LFI20:Rad:M}{18.7}
\setsymbol{LFI:knee:frequency:LFI21:Rad:M}{37.2}
\setsymbol{LFI:knee:frequency:LFI22:Rad:M}{12.7}
\setsymbol{LFI:knee:frequency:LFI23:Rad:M}{34.6}
\setsymbol{LFI:knee:frequency:LFI24:Rad:M}{46.2}
\setsymbol{LFI:knee:frequency:LFI25:Rad:M}{24.9}
\setsymbol{LFI:knee:frequency:LFI26:Rad:M}{67.6}
\setsymbol{LFI:knee:frequency:LFI27:Rad:M}{187.4}
\setsymbol{LFI:knee:frequency:LFI28:Rad:M}{122.2}
\setsymbol{LFI:knee:frequency:LFI18:Rad:S}{17.7}
\setsymbol{LFI:knee:frequency:LFI19:Rad:S}{22.0}
\setsymbol{LFI:knee:frequency:LFI20:Rad:S}{8.7}
\setsymbol{LFI:knee:frequency:LFI21:Rad:S}{25.9}
\setsymbol{LFI:knee:frequency:LFI22:Rad:S}{15.8}
\setsymbol{LFI:knee:frequency:LFI23:Rad:S}{129.8}
\setsymbol{LFI:knee:frequency:LFI24:Rad:S}{100.9}
\setsymbol{LFI:knee:frequency:LFI25:Rad:S}{38.9}
\setsymbol{LFI:knee:frequency:LFI26:Rad:S}{58.9}
\setsymbol{LFI:knee:frequency:LFI27:Rad:S}{104.4}
\setsymbol{LFI:knee:frequency:LFI28:Rad:S}{40.7}

\setsymbol{LFI:slope:70GHz:units}{$-1.03$\mHz}
\setsymbol{LFI:slope:44GHz:units}{$-0.89$\mHz}
\setsymbol{LFI:slope:30GHz:units}{$-0.87$\mHz}

\setsymbol{LFI:slope:70GHz}{$-1.03$}
\setsymbol{LFI:slope:44GHz}{$-0.89$}
\setsymbol{LFI:slope:30GHz}{$-0.87$}

\setsymbol{LFI:slope:LFI18:Rad:M:units}{$-1.04$\mHz}
\setsymbol{LFI:slope:LFI19:Rad:M:units}{$-1.09$\mHz}
\setsymbol{LFI:slope:LFI20:Rad:M:units}{$-0.69$\mHz}
\setsymbol{LFI:slope:LFI21:Rad:M:units}{$-1.56$\mHz}
\setsymbol{LFI:slope:LFI22:Rad:M:units}{$-1.01$\mHz}
\setsymbol{LFI:slope:LFI23:Rad:M:units}{$-0.96$\mHz}
\setsymbol{LFI:slope:LFI24:Rad:M:units}{$-0.83$\mHz}
\setsymbol{LFI:slope:LFI25:Rad:M:units}{$-0.91$\mHz}
\setsymbol{LFI:slope:LFI26:Rad:M:units}{$-0.95$\mHz}
\setsymbol{LFI:slope:LFI27:Rad:M:units}{$-0.87$\mHz}
\setsymbol{LFI:slope:LFI28:Rad:M:units}{$-0.88$\mHz}
\setsymbol{LFI:slope:LFI18:Rad:S:units}{$-1.15$\mHz}
\setsymbol{LFI:slope:LFI19:Rad:S:units}{$-1.00$\mHz}
\setsymbol{LFI:slope:LFI20:Rad:S:units}{$-0.95$\mHz}
\setsymbol{LFI:slope:LFI21:Rad:S:units}{$-0.92$\mHz}
\setsymbol{LFI:slope:LFI22:Rad:S:units}{$-1.01$\mHz}
\setsymbol{LFI:slope:LFI23:Rad:S:units}{$-0.95$\mHz}
\setsymbol{LFI:slope:LFI24:Rad:S:units}{$-0.73$\mHz}
\setsymbol{LFI:slope:LFI25:Rad:S:units}{$-1.16$\mHz}
\setsymbol{LFI:slope:LFI26:Rad:S:units}{$-0.79$\mHz}
\setsymbol{LFI:slope:LFI27:Rad:S:units}{$-0.82$\mHz}
\setsymbol{LFI:slope:LFI28:Rad:S:units}{$-0.91$\mHz}

\setsymbol{LFI:slope:LFI18:Rad:M}{$-1.04$}
\setsymbol{LFI:slope:LFI19:Rad:M}{$-1.09$}
\setsymbol{LFI:slope:LFI20:Rad:M}{$-0.69$}
\setsymbol{LFI:slope:LFI21:Rad:M}{$-1.56$}
\setsymbol{LFI:slope:LFI22:Rad:M}{$-1.01$}
\setsymbol{LFI:slope:LFI23:Rad:M}{$-0.96$}
\setsymbol{LFI:slope:LFI24:Rad:M}{$-0.83$}
\setsymbol{LFI:slope:LFI25:Rad:M}{$-0.91$}
\setsymbol{LFI:slope:LFI26:Rad:M}{$-0.95$}
\setsymbol{LFI:slope:LFI27:Rad:M}{$-0.87$}
\setsymbol{LFI:slope:LFI28:Rad:M}{$-0.88$}
\setsymbol{LFI:slope:LFI18:Rad:S}{$-1.15$}
\setsymbol{LFI:slope:LFI19:Rad:S}{$-1.00$}
\setsymbol{LFI:slope:LFI20:Rad:S}{$-0.95$}
\setsymbol{LFI:slope:LFI21:Rad:S}{$-0.92$}
\setsymbol{LFI:slope:LFI22:Rad:S}{$-1.01$}
\setsymbol{LFI:slope:LFI23:Rad:S}{$-0.95$}
\setsymbol{LFI:slope:LFI24:Rad:S}{$-0.73$}
\setsymbol{LFI:slope:LFI25:Rad:S}{$-1.16$}
\setsymbol{LFI:slope:LFI26:Rad:S}{$-0.79$}
\setsymbol{LFI:slope:LFI27:Rad:S}{$-0.82$}
\setsymbol{LFI:slope:LFI28:Rad:S}{$-0.91$}

\setsymbol{LFI:FWHM:70GHz:units}{13\parcm01}
\setsymbol{LFI:FWHM:44GHz:units}{27\parcm92}
\setsymbol{LFI:FWHM:30GHz:units}{32\parcm65}

\setsymbol{LFI:FWHM:70GHz}{13.01}
\setsymbol{LFI:FWHM:44GHz}{27.92}
\setsymbol{LFI:FWHM:30GHz}{32.65}

\setsymbol{LFI:FWHM:LFI18:units}{13\parcm39}
\setsymbol{LFI:FWHM:LFI19:units}{13\parcm01}
\setsymbol{LFI:FWHM:LFI20:units}{12\parcm75}
\setsymbol{LFI:FWHM:LFI21:units}{12\parcm74}
\setsymbol{LFI:FWHM:LFI22:units}{12\parcm87}
\setsymbol{LFI:FWHM:LFI23:units}{13\parcm27}
\setsymbol{LFI:FWHM:LFI24:units}{22\parcm98}
\setsymbol{LFI:FWHM:LFI25:units}{30\parcm46}
\setsymbol{LFI:FWHM:LFI26:units}{30\parcm31}
\setsymbol{LFI:FWHM:LFI27:units}{32\parcm65}
\setsymbol{LFI:FWHM:LFI28:units}{32\parcm66}

\setsymbol{LFI:FWHM:LFI18}{13.39}
\setsymbol{LFI:FWHM:LFI19}{13.01}
\setsymbol{LFI:FWHM:LFI20}{12.75}
\setsymbol{LFI:FWHM:LFI21}{12.74}
\setsymbol{LFI:FWHM:LFI22}{12.87}
\setsymbol{LFI:FWHM:LFI23}{13.27}
\setsymbol{LFI:FWHM:LFI24}{22.98}
\setsymbol{LFI:FWHM:LFI25}{30.46}
\setsymbol{LFI:FWHM:LFI26}{30.31}
\setsymbol{LFI:FWHM:LFI27}{32.65}
\setsymbol{LFI:FWHM:LFI28}{32.66}

\setsymbol{LFI:FWHM:uncertainty:LFI18:units}{0.170\arcm}
\setsymbol{LFI:FWHM:uncertainty:LFI19:units}{0.174\arcm}
\setsymbol{LFI:FWHM:uncertainty:LFI20:units}{0.170\arcm}
\setsymbol{LFI:FWHM:uncertainty:LFI21:units}{0.156\arcm}
\setsymbol{LFI:FWHM:uncertainty:LFI22:units}{0.164\arcm}
\setsymbol{LFI:FWHM:uncertainty:LFI23:units}{0.171\arcm}
\setsymbol{LFI:FWHM:uncertainty:LFI24:units}{0.652\arcm}
\setsymbol{LFI:FWHM:uncertainty:LFI25:units}{1.075\arcm}
\setsymbol{LFI:FWHM:uncertainty:LFI26:units}{1.131\arcm}
\setsymbol{LFI:FWHM:uncertainty:LFI27:units}{1.266\arcm}
\setsymbol{LFI:FWHM:uncertainty:LFI28:units}{1.287\arcm}

\setsymbol{LFI:FWHM:uncertainty:LFI18}{0.170}
\setsymbol{LFI:FWHM:uncertainty:LFI19}{0.174}
\setsymbol{LFI:FWHM:uncertainty:LFI20}{0.170}
\setsymbol{LFI:FWHM:uncertainty:LFI21}{0.156}
\setsymbol{LFI:FWHM:uncertainty:LFI22}{0.164}
\setsymbol{LFI:FWHM:uncertainty:LFI23}{0.171}
\setsymbol{LFI:FWHM:uncertainty:LFI24}{0.652}
\setsymbol{LFI:FWHM:uncertainty:LFI25}{1.075}
\setsymbol{LFI:FWHM:uncertainty:LFI26}{1.131}
\setsymbol{LFI:FWHM:uncertainty:LFI27}{1.266}
\setsymbol{LFI:FWHM:uncertainty:LFI28}{1.287}

\setsymbol{HFI:center:frequency:100GHz:units}{100\,GHz}
\setsymbol{HFI:center:frequency:143GHz:units}{143\,GHz}
\setsymbol{HFI:center:frequency:217GHz:units}{217\,GHz}
\setsymbol{HFI:center:frequency:353GHz:units}{353\,GHz}
\setsymbol{HFI:center:frequency:545GHz:units}{545\,GHz}
\setsymbol{HFI:center:frequency:857GHz:units}{857\,GHz}

\setsymbol{HFI:center:frequency:100GHz}{100}
\setsymbol{HFI:center:frequency:143GHz}{143}
\setsymbol{HFI:center:frequency:217GHz}{217}
\setsymbol{HFI:center:frequency:353GHz}{353}
\setsymbol{HFI:center:frequency:545GHz}{545}
\setsymbol{HFI:center:frequency:857GHz}{857}

\setsymbol{HFI:Ndetectors:100GHz}{8}
\setsymbol{HFI:Ndetectors:143GHz}{11}
\setsymbol{HFI:Ndetectors:217GHz}{12}
\setsymbol{HFI:Ndetectors:353GHz}{12}
\setsymbol{HFI:Ndetectors:545GHz}{3}
\setsymbol{HFI:Ndetectors:857GHz}{4}

\setsymbol{HFI:FWHM:Maps:100GHz:units}{9\parcm88}
\setsymbol{HFI:FWHM:Maps:143GHz:units}{7\parcm18}
\setsymbol{HFI:FWHM:Maps:217GHz:units}{4\parcm87}
\setsymbol{HFI:FWHM:Maps:353GHz:units}{4\parcm65}
\setsymbol{HFI:FWHM:Maps:545GHz:units}{4\parcm72}
\setsymbol{HFI:FWHM:Maps:857GHz:units}{4\parcm39}
\setsymbol{HFI:FWHM:Maps:100GHz}{9.88}
\setsymbol{HFI:FWHM:Maps:143GHz}{7.18}
\setsymbol{HFI:FWHM:Maps:217GHz}{4.87}
\setsymbol{HFI:FWHM:Maps:353GHz}{4.65}
\setsymbol{HFI:FWHM:Maps:545GHz}{4.72}
\setsymbol{HFI:FWHM:Maps:857GHz}{4.39}

\setsymbol{HFI:beam:ellipticity:Maps:100GHz}{1.15}
\setsymbol{HFI:beam:ellipticity:Maps:143GHz}{1.01}
\setsymbol{HFI:beam:ellipticity:Maps:217GHz}{1.06}
\setsymbol{HFI:beam:ellipticity:Maps:353GHz}{1.05}
\setsymbol{HFI:beam:ellipticity:Maps:545GHz}{1.14}
\setsymbol{HFI:beam:ellipticity:Maps:857GHz}{1.19}

\setsymbol{HFI:FWHM:Mars:100GHz:units}{9\parcm37}
\setsymbol{HFI:FWHM:Mars:143GHz:units}{7\parcm04}
\setsymbol{HFI:FWHM:Mars:217GHz:units}{4\parcm68}
\setsymbol{HFI:FWHM:Mars:353GHz:units}{4\parcm43}
\setsymbol{HFI:FWHM:Mars:545GHz:units}{3\parcm80}
\setsymbol{HFI:FWHM:Mars:857GHz:units}{3\parcm67}

\setsymbol{HFI:FWHM:Mars:100GHz}{9.37}
\setsymbol{HFI:FWHM:Mars:143GHz}{7.04}
\setsymbol{HFI:FWHM:Mars:217GHz}{4.68}
\setsymbol{HFI:FWHM:Mars:353GHz}{4.43}
\setsymbol{HFI:FWHM:Mars:545GHz}{3.80}
\setsymbol{HFI:FWHM:Mars:857GHz}{3.67}

\setsymbol{HFI:beam:ellipticity:Mars:100GHz}{1.18}
\setsymbol{HFI:beam:ellipticity:Mars:143GHz}{1.03}
\setsymbol{HFI:beam:ellipticity:Mars:217GHz}{1.14}
\setsymbol{HFI:beam:ellipticity:Mars:353GHz}{1.09}
\setsymbol{HFI:beam:ellipticity:Mars:545GHz}{1.25}
\setsymbol{HFI:beam:ellipticity:Mars:857GHz}{1.03}

\setsymbol{HFI:CMB:relative:calibration:100GHz}{$\lsim 1\%$}
\setsymbol{HFI:CMB:relative:calibration:143GHz}{$\lsim 1\%$}
\setsymbol{HFI:CMB:relative:calibration:217GHz}{$\lsim 1\%$}
\setsymbol{HFI:CMB:relative:calibration:353GHz}{$\lsim 1\%$}
\setsymbol{HFI:CMB:relative:calibration:545GHz}{}
\setsymbol{HFI:CMB:relative:calibration:857GHz}{}

\setsymbol{HFI:CMB:absolute:calibration:100GHz}{$\lsim 2\%$}
\setsymbol{HFI:CMB:absolute:calibration:143GHz}{$\lsim 2\%$}
\setsymbol{HFI:CMB:absolute:calibration:217GHz}{$\lsim 2\%$}
\setsymbol{HFI:CMB:absolute:calibration:353GHz}{$\lsim 2\%$}
\setsymbol{HFI:CMB:absolute:calibration:545GHz}{}
\setsymbol{HFI:CMB:absolute:calibration:857GHz}{}

\setsymbol{HFI:FIRAS:gain:calibration:accuracy:statistical:100GHz}{}
\setsymbol{HFI:FIRAS:gain:calibration:accuracy:statistical:143GHz}{}
\setsymbol{HFI:FIRAS:gain:calibration:accuracy:statistical:217GHz}{}
\setsymbol{HFI:FIRAS:gain:calibration:accuracy:statistical:353GHz}{2.5\%}
\setsymbol{HFI:FIRAS:gain:calibration:accuracy:statistical:545GHz}{1\%}
\setsymbol{HFI:FIRAS:gain:calibration:accuracy:statistical:857GHz}{0.5\%}

\setsymbol{HFI:FIRAS:gain:calibration:accuracy:systematic:100GHz}{}
\setsymbol{HFI:FIRAS:gain:calibration:accuracy:systematic:143GHz}{}
\setsymbol{HFI:FIRAS:gain:calibration:accuracy:systematic:217GHz}{}
\setsymbol{HFI:FIRAS:gain:calibration:accuracy:systematic:353GHz}{}
\setsymbol{HFI:FIRAS:gain:calibration:accuracy:systematic:545GHz}{7\%}
\setsymbol{HFI:FIRAS:gain:calibration:accuracy:systematic:857GHz}{7\%}

\setsymbol{HFI:FIRAS:zero:point:accuracy:100GHz:units}{0.8\MJysr}
\setsymbol{HFI:FIRAS:zero:point:accuracy:143GHz:units}{}
\setsymbol{HFI:FIRAS:zero:point:accuracy:217GHz:units}{}
\setsymbol{HFI:FIRAS:zero:point:accuracy:353GHz:units}{1.4\MJysr}
\setsymbol{HFI:FIRAS:zero:point:accuracy:545GHz:units}{2.2\MJysr}
\setsymbol{HFI:FIRAS:zero:point:accuracy:857GHz:units}{1.7\MJysr}

\setsymbol{HFI:FIRAS:zero:point:accuracy:100GHz}{0.8}
\setsymbol{HFI:FIRAS:zero:point:accuracy:143GHz}{}
\setsymbol{HFI:FIRAS:zero:point:accuracy:217GHz}{}
\setsymbol{HFI:FIRAS:zero:point:accuracy:353GHz}{1.4}
\setsymbol{HFI:FIRAS:zero:point:accuracy:545GHz}{2.2}
\setsymbol{HFI:FIRAS:zero:point:accuracy:857GHz}{1.7}

\setsymbol{HFI:unit:conversion:100GHz:units}{0.2415\MJysrmK}
\setsymbol{HFI:unit:conversion:143GHz:units}{0.3694\MJysrmK}
\setsymbol{HFI:unit:conversion:217GHz:units}{0.4811\MJysrmK}
\setsymbol{HFI:unit:conversion:353GHz:units}{0.2883\MJysrmK}
\setsymbol{HFI:unit:conversion:545GHz:units}{0.05826\MJysrmK}
\setsymbol{HFI:unit:conversion:857GHz:units}{0.002238\MJysrmK}

\setsymbol{HFI:unit:conversion:100GHz}{0.2415}
\setsymbol{HFI:unit:conversion:143GHz}{0.3694}
\setsymbol{HFI:unit:conversion:217GHz}{0.4811}
\setsymbol{HFI:unit:conversion:353GHz}{0.2883}
\setsymbol{HFI:unit:conversion:545GHz}{0.05826}
\setsymbol{HFI:unit:conversion:857GHz}{0.002238}

\setsymbol{HFI:colour:correction:alpha=-2:V1.01:100GHz}{0.9893}
\setsymbol{HFI:colour:correction:alpha=-2:V1.01:143GHz}{0.9759}
\setsymbol{HFI:colour:correction:alpha=-2:V1.01:217GHz}{1.0007}
\setsymbol{HFI:colour:correction:alpha=-2:V1.01:353GHz}{1.0028}
\setsymbol{HFI:colour:correction:alpha=-2:V1.01:545GHz}{1.0019}
\setsymbol{HFI:colour:correction:alpha=-2:V1.01:857GHz}{0.9889}

\setsymbol{HFI:colour:correction:alpha=0:V1.01:100GHz}{1.0008}
\setsymbol{HFI:colour:correction:alpha=0:V1.01:143GHz}{1.0148}
\setsymbol{HFI:colour:correction:alpha=0:V1.01:217GHz}{0.9909}
\setsymbol{HFI:colour:correction:alpha=0:V1.01:353GHz}{0.9888}
\setsymbol{HFI:colour:correction:alpha=0:V1.01:545GHz}{0.9878}
\setsymbol{HFI:colour:correction:alpha=0:V1.01:857GHz}{1.0014}

\providecommand{\sorthelp}[1]{}

\usepackage{natbib}
\usepackage{hyperref}
\usepackage{breakurl}
\usepackage{graphicx}
\usepackage{fixltx2e}
\usepackage{txfonts}

\newcommand{\ha}{H$\alpha$} 
\newcommand{\hi}{\ion{H}{i}} 
\newcommand{\hii}{\ion{H}{ii}} 

\begin{document}
%This author list corresponds to \title{Author list for SVN PIP\_96\_Proj\_7\_2\_Alves: Dust emission at millimetre wavelengths in the Galactic plane}
%Prepared by R. Leonardi (rleonardi@sciops.esa.int), ESAC/ESA
%This version is from Wed Jan 15 15:11:07 2014 CET
%\subtitle{There are 194 co-authors in this list}
\author{\small
Planck Collaboration:
P.~A.~R.~Ade\inst{78}
\and
N.~Aghanim\inst{54}
\and
M.~I.~R.~Alves\inst{54}\thanks{Corresponding author: M. I. R. Alves \url{marta.alves@ias.u-psud.fr}}
\and
M.~Arnaud\inst{67}
\and
M.~Ashdown\inst{64, 6}
\and
F.~Atrio-Barandela\inst{18}
\and
J.~Aumont\inst{54}
\and
C.~Baccigalupi\inst{77}
\and
A.~J.~Banday\inst{82, 10}
% ...
}
\institute{\small
APC, AstroParticule et Cosmologie, Universit\'{e} Paris Diderot, CNRS/IN2P3, CEA/lrfu, Observatoire de Paris, Sorbonne Paris Cit\'{e}, 10, rue Alice Domon et L\'{e}onie Duquet, 75205 Paris Cedex 13, France\\
\and
Aalto University Mets\"{a}hovi Radio Observatory and Dept of Radio Science and Engineering, P.O. Box 13000, FI-00076 AALTO, Finland\\
\and
African Institute for Mathematical Sciences, 6-8 Melrose Road, Muizenberg, Cape Town, South Africa\\
\and
Agenzia Spaziale Italiana Science Data Center, Via del Politecnico snc, 00133, Roma, Italy\\
\and
Agenzia Spaziale Italiana, Viale Liegi 26, Roma, Italy\\
\and
Astrophysics Group, Cavendish Laboratory, University of Cambridge, J J Thomson Avenue, Cambridge CB3 0HE, U.K.\\
\and
Astrophysics \& Cosmology Research Unit, School of Mathematics, Statistics \& Computer Science, University of KwaZulu-Natal, Westville Campus, Private Bag X54001, Durban 4000, South Africa\\
\and
Atacama Large Millimeter/submillimeter Array, ALMA Santiago Central Offices, Alonso de Cordova 3107, Vitacura, Casilla 763 0355, Santiago, Chile\\
\and
CITA, University of Toronto, 60 St. George St., Toronto, ON M5S 3H8, Canada\\
\and
CNRS, IRAP, 9 Av. colonel Roche, BP 44346, F-31028 Toulouse cedex 4, France\\
\and
California Institute of Technology, Pasadena, California, U.S.A.\\
\and
Centro de Estudios de F\'{i}sica del Cosmos de Arag\'{o}n (CEFCA), Plaza San Juan, 1, planta 2, E-44001, Teruel, Spain\\
\and
Computational Cosmology Center, Lawrence Berkeley National Laboratory, Berkeley, California, U.S.A.\\
\and
Consejo Superior de Investigaciones Cient\'{\i}ficas (CSIC), Madrid, Spain\\
\and
DSM/Irfu/SPP, CEA-Saclay, F-91191 Gif-sur-Yvette Cedex, France\\
\and
DTU Space, National Space Institute, Technical University of Denmark, Elektrovej 327, DK-2800 Kgs. Lyngby, Denmark\\
\and
D\'{e}partement de Physique Th\'{e}orique, Universit\'{e} de Gen\`{e}ve, 24, Quai E. Ansermet,1211 Gen\`{e}ve 4, Switzerland\\
\and
Departamento de F\'{\i}sica Fundamental, Facultad de Ciencias, Universidad de Salamanca, 37008 Salamanca, Spain\\
\and
Departamento de F\'{\i}sica, Universidad de Oviedo, Avda. Calvo Sotelo s/n, Oviedo, Spain\\
\and
Department of Astrophysics/IMAPP, Radboud University Nijmegen, P.O. Box 9010, 6500 GL Nijmegen, The Netherlands\\
\and
Department of Electrical Engineering and Computer Sciences, University of California, Berkeley, California, U.S.A.\\
\and
Department of Physics \& Astronomy, University of British Columbia, 6224 Agricultural Road, Vancouver, British Columbia, Canada\\
\and
Department of Physics and Astronomy, Dana and David Dornsife College of Letter, Arts and Sciences, University of Southern California, Los Angeles, CA 90089, U.S.A.\\
\and
Department of Physics and Astronomy, University College London, London WC1E 6BT, U.K.\\
\and
Department of Physics, Gustaf H\"{a}llstr\"{o}min katu 2a, University of Helsinki, Helsinki, Finland\\
\and
Department of Physics, Princeton University, Princeton, New Jersey, U.S.A.\\
\and
Department of Physics, University of California, Santa Barbara, California, U.S.A.\\
\and
Department of Physics, University of Illinois at Urbana-Champaign, 1110 West Green Street, Urbana, Illinois, U.S.A.\\
\and
Dipartimento di Fisica e Astronomia G. Galilei, Universit\`{a} degli Studi di Padova, via Marzolo 8, 35131 Padova, Italy\\
\and
Dipartimento di Fisica e Scienze della Terra, Universit\`{a} di Ferrara, Via Saragat 1, 44122 Ferrara, Italy\\
\and
Dipartimento di Fisica, Universit\`{a} La Sapienza, P. le A. Moro 2, Roma, Italy\\
\and
Dipartimento di Fisica, Universit\`{a} degli Studi di Milano, Via Celoria, 16, Milano, Italy\\
\and
Dipartimento di Fisica, Universit\`{a} degli Studi di Trieste, via A. Valerio 2, Trieste, Italy\\
\and
Dipartimento di Fisica, Universit\`{a} di Roma Tor Vergata, Via della Ricerca Scientifica, 1, Roma, Italy\\
\and
Discovery Center, Niels Bohr Institute, Blegdamsvej 17, Copenhagen, Denmark\\
\and
Dpto. Astrof\'{i}sica, Universidad de La Laguna (ULL), E-38206 La Laguna, Tenerife, Spain\\
\and
European Southern Observatory, ESO Vitacura, Alonso de Cordova 3107, Vitacura, Casilla 19001, Santiago, Chile\\
\and
European Space Agency, ESAC, Planck Science Office, Camino bajo del Castillo, s/n, Urbanizaci\'{o}n Villafranca del Castillo, Villanueva de la Ca\~{n}ada, Madrid, Spain\\
\and
European Space Agency, ESTEC, Keplerlaan 1, 2201 AZ Noordwijk, The Netherlands\\
\and
Helsinki Institute of Physics, Gustaf H\"{a}llstr\"{o}min katu 2, University of Helsinki, Helsinki, Finland\\
\and
INAF - Osservatorio Astronomico di Padova, Vicolo dell'Osservatorio 5, Padova, Italy\\
\and
INAF - Osservatorio Astronomico di Roma, via di Frascati 33, Monte Porzio Catone, Italy\\
\and
INAF - Osservatorio Astronomico di Trieste, Via G.B. Tiepolo 11, Trieste, Italy\\
\and
INAF/IASF Bologna, Via Gobetti 101, Bologna, Italy\\
\and
INAF/IASF Milano, Via E. Bassini 15, Milano, Italy\\
\and
INFN, Sezione di Bologna, Via Irnerio 46, I-40126, Bologna, Italy\\
\and
INFN, Sezione di Roma 1, Universit\`{a} di Roma Sapienza, Piazzale Aldo Moro 2, 00185, Roma, Italy\\
\and
INFN/National Institute for Nuclear Physics, Via Valerio 2, I-34127 Trieste, Italy\\
\and
IPAG: Institut de Plan\'{e}tologie et d'Astrophysique de Grenoble, Universit\'{e} Joseph Fourier, Grenoble 1 / CNRS-INSU, UMR 5274, Grenoble, F-38041, France\\
\and
IUCAA, Post Bag 4, Ganeshkhind, Pune University Campus, Pune 411 007, India\\
\and
Imperial College London, Astrophysics group, Blackett Laboratory, Prince Consort Road, London, SW7 2AZ, U.K.\\
\and
Infrared Processing and Analysis Center, California Institute of Technology, Pasadena, CA 91125, U.S.A.\\
\and
Institut Universitaire de France, 103, bd Saint-Michel, 75005, Paris, France\\
\and
Institut d'Astrophysique Spatiale, CNRS (UMR8617) Universit\'{e} Paris-Sud 11, B\^{a}timent 121, Orsay, France\\
\and
Institut d'Astrophysique de Paris, CNRS (UMR7095), 98 bis Boulevard Arago, F-75014, Paris, France\\
\and
Institute for Space Sciences, Bucharest-Magurale, Romania\\
\and
Institute of Astronomy and Astrophysics, Academia Sinica, Taipei, Taiwan\\
\and
Institute of Astronomy, University of Cambridge, Madingley Road, Cambridge CB3 0HA, U.K.\\
\and
Institute of Theoretical Astrophysics, University of Oslo, Blindern, Oslo, Norway\\
\and
Instituto de Astrof\'{\i}sica de Canarias, C/V\'{\i}a L\'{a}ctea s/n, La Laguna, Tenerife, Spain\\
\and
Instituto de F\'{\i}sica de Cantabria (CSIC-Universidad de Cantabria), Avda. de los Castros s/n, Santander, Spain\\
\and
Jet Propulsion Laboratory, California Institute of Technology, 4800 Oak Grove Drive, Pasadena, California, U.S.A.\\
\and
Jodrell Bank Centre for Astrophysics, Alan Turing Building, School of Physics and Astronomy, The University of Manchester, Oxford Road, Manchester, M13 9PL, U.K.\\
\and
Kavli Institute for Cosmology Cambridge, Madingley Road, Cambridge, CB3 0HA, U.K.\\
\and
LAL, Universit\'{e} Paris-Sud, CNRS/IN2P3, Orsay, France\\
\and
LERMA, CNRS, Observatoire de Paris, 61 Avenue de l'Observatoire, Paris, France\\
\and
Laboratoire AIM, IRFU/Service d'Astrophysique - CEA/DSM - CNRS - Universit\'{e} Paris Diderot, B\^{a}t. 709, CEA-Saclay, F-91191 Gif-sur-Yvette Cedex, France\\
\and
Laboratoire Traitement et Communication de l'Information, CNRS (UMR 5141) and T\'{e}l\'{e}com ParisTech, 46 rue Barrault F-75634 Paris Cedex 13, France\\
\and
Laboratoire de Physique Subatomique et de Cosmologie, Universit\'{e} Joseph Fourier Grenoble I, CNRS/IN2P3, Institut National Polytechnique de Grenoble, 53 rue des Martyrs, 38026 Grenoble cedex, France\\
\and
Laboratoire de Physique Th\'{e}orique, Universit\'{e} Paris-Sud 11 \& CNRS, B\^{a}timent 210, 91405 Orsay, France\\
\and
Lawrence Berkeley National Laboratory, Berkeley, California, U.S.A.\\
\and
Max-Planck-Institut f\"{u}r Astrophysik, Karl-Schwarzschild-Str. 1, 85741 Garching, Germany\\
\and
National University of Ireland, Department of Experimental Physics, Maynooth, Co. Kildare, Ireland\\
\and
Niels Bohr Institute, Blegdamsvej 17, Copenhagen, Denmark\\
\and
Observational Cosmology, Mail Stop 367-17, California Institute of Technology, Pasadena, CA, 91125, U.S.A.\\
\and
Optical Science Laboratory, University College London, Gower Street, London, U.K.\\
\and
SISSA, Astrophysics Sector, via Bonomea 265, 34136, Trieste, Italy\\
\and
School of Physics and Astronomy, Cardiff University, Queens Buildings, The Parade, Cardiff, CF24 3AA, U.K.\\
\and
Space Sciences Laboratory, University of California, Berkeley, California, U.S.A.\\
\and
Special Astrophysical Observatory, Russian Academy of Sciences, Nizhnij Arkhyz, Zelenchukskiy region, Karachai-Cherkessian Republic, 369167, Russia\\
\and
UPMC Univ Paris 06, UMR7095, 98 bis Boulevard Arago, F-75014, Paris, France\\
\and
Universit\'{e} de Toulouse, UPS-OMP, IRAP, F-31028 Toulouse cedex 4, France\\
% ...
}

\title{\textit{Planck} intermediate results. XIV. Dust emission at millimetre
     wavelengths in the Galactic plane }

   \date{Received 25 July 2013/ Accepted 28 January 2014}

  \abstract
  {We use \Planck~HFI data combined with ancillary radio data to
    study the emissivity index of the interstellar dust emission in the
    frequency range 100--353\,GHz, or 3--0.8\,mm, in the Galactic
    plane. We analyse the region $l=20\degr$--$44\degr$ and $|b| \leq
    4\degr$ where the free-free emission can be estimated from radio
    recombination line data. We fit the spectra at each sky pixel with a
    modified blackbody model and two opacity spectral indices, $\beta_{\rm mm}$
    and $\beta_{\rm FIR}$, below and above 353\,GHz, respectively.
    We find that $\beta_{\rm mm}$ is
    smaller than $\beta_{\rm FIR}$, and we detect a
    correlation between this low frequency power-law index and the
    dust optical depth at 353\,GHz, $\tau_{353}$. The opacity spectral
    index $\beta_{\rm mm}$ increases from
    about 1.54 in the more diffuse regions of the Galactic disk, $|b|
    = 3\degr$--4$\degr$ and $\tau_{353} \sim 5\times 10^{-5}$, to
    about 1.66 in the densest regions with an optical depth of more than one
    order of magnitude higher. We associate
    this correlation with an evolution of the dust emissivity related to the
    fraction of molecular gas along the line of sight. This translates
    into $\beta_{\rm mm} \sim 1.54$ for a medium that is mostly atomic
    and $\beta_{\rm mm} \sim 1.66$ when the medium is dominated by
    molecular gas. 
  We find that both the two-level system model and magnetic dipole emission by
    ferromagnetic particles can explain the results.
   These results improve our understanding of the physics of
    interstellar dust and lead towards a complete model of the dust
    spectrum of the Milky Way from far-infrared to millimetre wavelengths.

 }
   {}

   \keywords{ISM: general -- Galaxy: general -- radiation mechanisms: general -- radio continuum: ISM -- submillimeter: ISM}

\titlerunning{Dust emission at millimetre
     wavelengths in the Galactic plane}

\authorrunning{Planck Collaboration}
\maketitle

%________________________________________________________________
%________________________________________________________________

\section{Introduction}
\label{sec:intro}

The frequency coverage of
\Planck\footnote{\Planck\ (\url{http://www.esa.int/Planck}) is a
  project of the European Space Agency (ESA) with instruments
  provided by two scientific consortia funded by ESA member states (in
  particular the lead countries France and Italy), with contributions
  from NASA (USA) and telescope reflectors provided by a collaboration
  between ESA and a scientific consortium led and funded by Denmark.}
is opening new windows in our understanding of Galactic emission. This
is especially the case for the high frequency data that provide an all-sky view of the
Rayleigh-Jeans regime of the thermal dust spectrum. %, until now unavailable at less than 10 arcmin resolution.
The emission at the \Planck~high frequency bands (350\,\microns--3\,mm)
is dominated by the contribution of big grains (radius larger
  than 0.05\,\microns), which heated by
stellar photons are in thermal equilibrium with the interstellar
radiation field (ISRF). The spectral energy distribution (SED) of big dust grains is usually approximated by a modified blackbody emission law of the form
\begin{equation}
I_{\nu} = \tau_{\nu_{0}} \left(\frac{\nu}{\nu_{0}} \right)^{\beta}
B_{\nu}(\nu,T_{\rm d}),
\label{eq:graybody}
\end{equation}
where $\tau_{\nu_{0}} $ is the dust optical depth at a reference
frequency $\nu_{0}$,
$\beta$ the spectral index of the opacity, and $B_{\nu}$ the Planck function, which depends on
both the frequency and the dust temperature $T_{\rm d}$. 
However, early
observations by the \textit{Cosmic Background Explorer} (\textit{COBE})
indicated that the modified blackbody spectrum does not provide a
good description of the dust SED from far-infrared (FIR) to
millimetre wavelengths \citep{Reach:1995}. Later works have confirmed
that $\beta$ appears to vary with frequency, the SED flattening in the
millimetre relative to the best single modified blackbody fit and
also varying with environment \citep{Finkbeiner:1999, Galliano:2005,
  Paladini:2007, planck2011-6.4b, planck2011-7.0}. 
Studies of dust analogues (e.g., \citealt{Agladze:1996}, \citealt{Boudet:2005}, \citealt{Coupeaud:2011}) have characterized the FIR and
millimetre emission of different types of amorphous silicates. These
show a frequency, as well as temperature, dependence of $\beta$ not
unlike the astronomical results. The astrophysical
interpretation of this flattening is under study 
as new observations
become available, and some possible explanations have been
suggested. 
One possibility is a description of the opacity of the big
grains in terms of a two-level system (TLS,
\citealt{Meny:2007}). Alternatively, it might be attributed to
magnetic dipole emission from magnetic particles \citep{DraineHensley:2013} or to the evolution
of carbon dust \citep{Jones:2013}.

To study the low frequency flattening of the dust SED in the Galactic plane, we need to
take the free-free emission from the ionized gas into account.
Free-free emission is a principal foreground contaminant of the cosmic microwave background (CMB), not only at radio frequencies, where it is comparable to other Galactic components such as synchrotron, but also at millimetre wavelengths where the thermal dust emission dominates. It becomes a major component in the Galactic plane where it is produced by the gas ionized by recently formed massive stars.
All-sky maps of the free-free emission, derived in the context of CMB
foreground studies, have been obtained directly from \ha~measurements
\citep{D3:2003, Finkbeiner:2003}. However, this optical line suffers
from large dust absorption along the Galactic plane, and thus fails to provide a reliable measure of the thermal emission at low Galactic latitudes.
A free-free map that includes the Galactic plane is essential, not only
to correctly evaluate the CMB power spectrum at low angular frequency,
but also for Galactic star formation studies. 
The \textit{WMAP} satellite has provided all-sky maps at five
microwave frequencies that have been combined to estimate the
contribution of free-free, synchrotron, thermal dust and anomalous
microwave emission (AME) using
a maximum entropy method (MEM, \citealt{Bennett:2013:WMAP9}). Another approach using hydrogen radio recombination lines (RRLs) has
been presented recently by \citet{Alves:2010,Alves:2012}. In contrast
to \ha, these radio lines at a frequency of 1.4\,GHz are optically thin
and are not absorbed by dust or the radio emitting plasma. The RRL method has provided the first direct measure of the diffuse free-free emission along the Galactic plane, in the longitude range $l = 20\degr$--$44\degr$ and for latitudes $|b| \leq 4\degr$ \citep{Alves:2012}.

The free-free emission has a spectral index $\alpha \equiv
  d\ln(I_{\nu})/d\ln\nu$, varying from $-0.10$ at 1.4\,GHz to
  $-0.15$ at 100\,GHz.
The free-free emission dominates at frequencies between
60 and 100\,GHz, but there the other Galactic components, namely
synchrotron, dust and AME,
also contribute to the total intensity. The AME is an additional Galactic
component observed in the frequency range 10--60\,GHz 
(e.g., \citealt{Kogut:1996}, \citealt{Leitch:1997},
\citealt{deOliveira-costa:1997}, \citealt{planck2011-7.2}, \citealt{planck2013-XII}, \citealt{planck2013-XV}) which cannot be explained by
free-free, synchrotron or thermal dust emission and is thought to
arise from small spinning dust grains \citep{Draine:1998a,Ali-Haimoud:2009,Ysard:2010a,Hoang:2010,Hoang:2011}. 
In a spectral decomposition of the four Galactic emission components
along the plane, Planck Collaboration Int. XXII (in prep.) find that the AME contribution is comparable to that of the
free-free in the frequency range 20--40\,GHz. 
On the other hand, and due to its steeper spectral index, $ -1.2
\la \alpha \la -0.7
$ \citep{Davies:1996,Ghosh:2012}, the synchrotron emission is
mostly dominant at frequencies less than a few gigahertz. 

This paper aims to characterize for the first time the dust emissivity in
the frequency range 100--353\,GHz of the diffuse emission in the
Galactic plane. For this purpose, we remove the free-free emission
contribution using the RRL data (Sect.~\ref{subsec:rrldata}). 
We start by describing the \Planck~and
ancillary data used in this work in Sects. \ref{sec:planckdata} and
\ref{sec:ancdata}. In Sect.~\ref{sec:analysis1} we present the data
analysis techniques, followed by the main results of the paper in
Sect.~\ref{sec:results}. These are discussed further and interpreted
in Sect.~\ref{sec:interpretation}, followed by the conclusions in
Sect.~\ref{sec:conc}.

%________________________________________________________________
%________________________________________________________________

\section{\Planck~HFI data}
\label{sec:planckdata}

\Planck~\citep{tauber2010a, planck2011-1.1,planck2013-p01} is the third-generation
space mission to measure the anisotropy of the CMB. It observed the sky in nine frequency bands covering
28.5--857\,GHz with high sensitivity and angular resolution from
$32\parcm24$ to $4\parcm33$. The Low Frequency Instrument (LFI;
\citealt{Mandolesi2010, Bersanelli2010, planck2011-1.4, planck2013-p02}) covered the
28.4, 44.1, and 70.4\,GHz bands with amplifiers cooled to 20\,K. The
High Frequency Instrument (HFI; \citealt{Lamarre2010, planck2011-1.5, planck2013-p03})
covered the 100, 143, 217, 353, 545, and 857\,GHz bands with bolometers
cooled to 0.1\,K. Polarization is measured in all but the
highest two bands \citep{Leahy2010, Rosset2010}. 

In the present work we use data from the \Planck\ 2013 data release which can be obtained from the
\Planck~Legacy Archive\footnote{\url{http://www.sciops.esa.int/index.php?project=planck&page=Planck_Legacy_Archive}}. We use the HFI 
data acquired between 13 August 2009 and 27 November
2010. These are converted to intensity units of
  MJy\,sr\mo~following the \textit{IRAS} SED convention
  \citep{Beichman:1988}, which assumes a spectral index
  $\alpha=-1$. Colour corrections based on the observed emission
  spectrum and on the spectral response of the receiver, are applied
  to derive the specific intensity at the effective frequency of each
  band \citep{planck2013-p03d}. The Planck CMB map is derived from the {\it SMICA} component
separation method and presented in \citet{planck2013-p06}. Close to the
plane of the Galaxy, it is not possible to correctly separate the CMB
fluctuations from the much brighter Galactic emission. Hence, this
region of the {\it SMICA} map has been replaced by a constrained
realization of the CMB fluctuations. For this reason we derive our
results using the data uncorrected for the CMB fluctuations, which we compare with those obtained when the {\it SMICA}
CMB map is subtracted from the data (Sect. \ref{subsec:smicacmb}). The
lowest HFI frequency band
also has the lowest angular resolution, of $9\parcm65$. However we
smooth the \Planck~data to a common resolution of 15\arcm, assuming Gaussian beams, to
match the lower resolution of the RRL data (Sect.~\ref{subsec:rrldata}). 

The 100\,GHz data are significantly contaminated by the CO $J$=1$\rightarrow$0 line at 115\,GHz and the 217\,GHz data by the  CO $J$=2$\rightarrow$1 line at 230\,GHz. 
At 353\,GHz the
contribution of the CO $J$=3$\rightarrow$2 line is small, but not negligible compared to dust
emission.
The CO line emission is subtracted using the \Planck~{\sc type} 1 CO maps from the MILCA (Modified
Independent Linear Combination Algorithm, \citealt{Hurier:2013})
bolometer solution \citep{planck2013-p03a}. These are converted from line integrated units to intensity units as described in
\citet{planck2013-p03a}. The calibration uncertainties on these maps are of
10, 2, and 5\,\% at 100, 217, and 353\,GHz, respectively. 
The 100\,GHz MILCA map has been compared with ground-based data, in
particular the \citet{Dame:2001} $^{12}$CO $J$=1$\rightarrow$0 survey along
the Galactic plane, for which there is an overall
agreement of 16\,\% \citep{planck2013-p03a}. However, in the Galactic plane region
of the present study, both datasets agree within 25\,\%. This
discrepancy can be explained by the shifting of the CO line frequency due
to Doppler effects, that is to say, the rotation of the Galactic disk \citep{planck2013-p03a}.

The overall calibration uncertainties for the \Planck~HFI maps are
10\,\% at 857 and 545\,GHz, 1.2\,\% at 353\,GHz, and 0.5\,\% at lower frequencies.  
These values are increased at the lowest frequencies due to the
subtraction of the CO and free-free emission.
We did not subtract the zodiacal dust emission from the maps, because it is a
negligible contribution in the Galactic plane
\citep{planck2013-pip88}. Moreover, the
cosmic infrared background (CIB) monopole was removed from all the HFI
maps as described in \citet{planck2013-p06b}.

%________________________________________________________________
%________________________________________________________________

\section{Ancillary data}
\label{sec:ancdata}

Along with \Planck~HFI we need to use ancillary data, namely RRL
observations for the removal of the free-free emission and \textit{IRAS} data
to constrain the dust temperature. 
All data sets are in
{\tt HEALPix} format \citep{Gorski:2005}, at N$_{\rm side}=512$, and are smoothed to a common resolution of 15\arcm.

%________________________________________________________________
\subsection{Radio Recombination Line data}
\label{subsec:rrldata}

A fully-sampled map of the free-free emission in the Galactic plane
region $l = 20\degr$--$44\degr$ and $|b| \leq 4\degr$ has been
derived by \citet{Alves:2012} using RRL data. These data are from
the \hi~Parkes All-Sky Survey and associated Zone of Avoidance Survey \citep{Staveley-Smith:1996,Staveley-Smith:1998}
 at 1.4\,GHz and 15\arcm~resolution. One source of
   uncertainty on these data is the conversion from the observed antenna
temperature to intensity units, which requires a detailed knowledge of
the observing beam \citep{Rohlfs:2000}. The RRL data presented in \citeauthor{Alves:2012} may need a correction downwards of
 5–-10\,\%, since they were converted to a
 scale appropriate for point sources; this correction depends on the
 angular size of the source relative to the main beam of 15\arcm. The free-free brightness
   temperature\footnote{Following the definition of brightness
     temperature by \citet{Spitzer:1978}, in the Rayleigh-Jeans regime.} estimated from the
   RRL integrated line emission depends on the electron temperature of
   the ionized gas as $T_{\rm e}^{1.15}$ \citep{Gordon:2009}. \citeauthor{Alves:2012} used an average value of
 $T_{\rm e}=6000$\,K; an increase of this value by 500 \,K
 (1000\,K) would increase the brightness temperature by 10\,\%
 (19\,\%). 

The RRL free-free data are similarly used in the work of Planck Collaboration Int. XXII (in prep.) to separate the different emission
components in the Galactic plane and to determine the contribution
of the AME. In that work, the free-free map estimated from the radio data are compared to two other
free-free solutions, given by the \Planck\ {\tt fastMEM} (Planck Collaboration Int. XXII in prep.) and
\textit{WMAP} MEM \citep{Bennett:2013:WMAP9} component separation
methods. The {\tt fastMEM} and \textit{WMAP} results agree within 2\,\% but
they are about 20\,\% higher than the RRL estimation. The proposed
solution to this difference is to scale the free-free map from
\citet{Alves:2012} upwards by 10\,\%, which is equivalent to increasing the electron temperature to
7000\,K (Planck Collaboration Int. XXII in prep.). In this paper we use the same electron
temperature of 7000\,K and adopt an overall calibration uncertainty of
10\,\% in the free-free continuum estimated from the RRL data.

%________________________________________________________________
\subsection{IRAS data}
\label{subsec:irasdata}

We use the IRIS (Improved Reprocessing of the \textit{IRAS} survey) data at
100\microns~\citep{Miville-Deschenes:2005} to constrain the peak of the
thermal dust emission. The calibration uncertainty for these data is 13.5\,\%.

%________________________________________________________________

\subsection{\hi~data}
\label{subsec:hidata}

The \hi~data from the Galactic All-Sky Survey (GASS,
\citealt{McClure-Griffiths:2009}) are used to estimate the column
density of the atomic medium. The GASS survey
mapped the 21-cm line emission in the southern sky, $\delta < 1\degr$,
at $14\farcm4$ angular resolution and 1\kms~velocity resolution. We
use the data corrected for instrumental effects, stray radiation and radio frequency
interference from \citet{Kalberla:2010}. The average
temperature uncertainties for these data are below 1\,\%. The \hi~line is integrated as
described in \citet{planck2013-XVII} and converted to \hi~column density
assuming that the line is optically thin. The optically thin limit is
a simplistic approach in the Galactic plane and results in an
underestimation of the true column density, by about 30--50\,\% as
found in \hi~continuum absorption studies \citep{Strasser:2004}.

%________________________________________________________________
%________________________________________________________________
\section{Analysis}
\label{sec:analysis1}

The aim here is to determine the power-law index of the interstellar
dust opacity at the lowest HFI frequencies, which we do by fitting the dust SED.

As mentioned in Sect.~\ref{sec:intro}, $\beta$ appears to be
frequency dependent with a break observed at frequencies around
600\,GHz, or $\lambda \sim 500\micron$ \citep{Paradis:2009,Gordon:2010,Galliano:2011}. \citet{planck2011-7.13}
also found that a single modified blackbody curve accurately fits the
FIR spectrum
of Galactic molecular clouds, but leaves large residuals at frequencies below 353\,GHz.
For this reason we
decided to fit the dust SED using a modified blackbody model, but
allowing $\beta$ to vary with frequency, having $\beta=\beta_{\rm FIR}$ for $\nu \geq 353$\,GHz and $\beta=\beta_{\rm mm}$ for
$\nu < 353$\,GHz. Using the \Planck~HFI bands along with the \textit{IRAS} 100\micron~data, we also solve for the other parameters in Eq. (\ref{eq:graybody}), namely $T_{\rm d}$ and $\tau_{353}$, where we take
the reference frequency as 353\,GHz for the dust optical depth.

The \Planck~maps at frequencies above 353\,GHz contain mainly dust emission and
also
CIB emission. The CIB fluctuations
have a power spectrum flatter than that of the interstellar dust
\citep{Miville-Deschenes:2002, Lagache:2007, planck2011-6.6}, thus
contributing mostly at small angular scales and producing a statistically
homogeneous signal. This signal only represents a significant fraction
of the total brightness in the most diffuse high latitude regions of
the sky, and thus can be neglected in the Galactic plane.

In the range 100--353\,GHz, even though most of the emission comes from interstellar dust, both the CMB and
free-free components also contribute to the total brightness. 
The fluctuations of the CMB are faint, rms of about 80\,$\mu$K at 
scales of 15\arcm, compared to the brightest emission in the
Galactic plane. Therefore, we can neglect the contribution from the
CMB fluctuations, since its rms temperature is about 5\,\% of the total emission
in the thin disk of the Galaxy. However, at latitudes $|b| \ga
2\degr$ the CMB fluctuations at 100\,GHz are about 10 times
brighter than the free-free emission. The effects of neglecting the CMB component at these higher
latitudes will be investigated via simulations in Sect.
\ref{subsec:simulations}, as well as using the {\it SMICA} CMB map in
Sect.~\ref{subsec:smicacmb}. 

\begin{figure}
\hspace{-0.4cm}
\includegraphics[scale=0.36,angle=90]{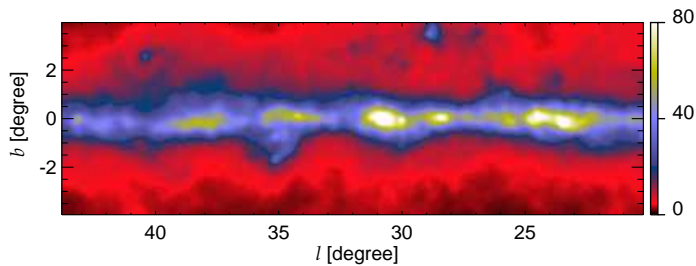}
\vspace{0.1cm}
\caption{HFI 353\,GHz map of the Galactic plane region $l=20\degr$--$44\degr$, $|b| \leq 4\degr$, in units of \MJysr~and at 15\arcm~resolution.}
\label{fig:maps}
\end{figure}

At $|b| \la 1\degr$, the contribution of the free-free emission
can be as high as 20--40\,\% to the total
emission at 100\,GHz, from both the diffuse and the individual \hii~regions.
Therefore, we need to remove the free-free emission if we are to fit the dust
spectrum only with a modified blackbody model.
For this purpose, we use the free-free map estimated from the RRLs
(Sect.~\ref{subsec:rrldata}), as
this is currently the only direct measure of this emission in
the Galactic plane, in particular in the $24\degr \times 8\degr$ region centred on $(l{,}b) = (32\degr{,}
0\degr)$. This region is shown in Fig.~\ref{fig:maps}, in the
\Planck~353\,GHz channel. The free-free continuum, estimated
from the RRL data at 1.4\,GHz, is extrapolated to the HFI frequencies using a frequency dependent Gaunt factor (Eq. 10.9
of \citealt{Draine:2011}) and an electron temperature of 7000\,K.

We used the {\tt IDL} {\tt MPFIT} routine to fit the final SEDs pixel-by-pixel in
the $24\degr \times 8\degr$ region. This routine performs weighted least-squares
fitting of the data \citep{Markwardt:2009}, taking into account the
noise (both statistical noise and systematic uncertainties)
for each spectral band. 
We also include a noise term from the CMB fluctuations,
typically 80\,$\mu$K, which will be dominant outside the Galactic plane
and at the lowest frequencies.
These uncertainties are used to give weights to the
spectral points.
Colour corrections based on the local spectral index across each band were applied to both \Planck\ and \textit{IRAS} data during the
model-fitting procedure \citep{planck2013-p03d}.

\begin{figure}
\centering
\includegraphics[width=1\columnwidth]{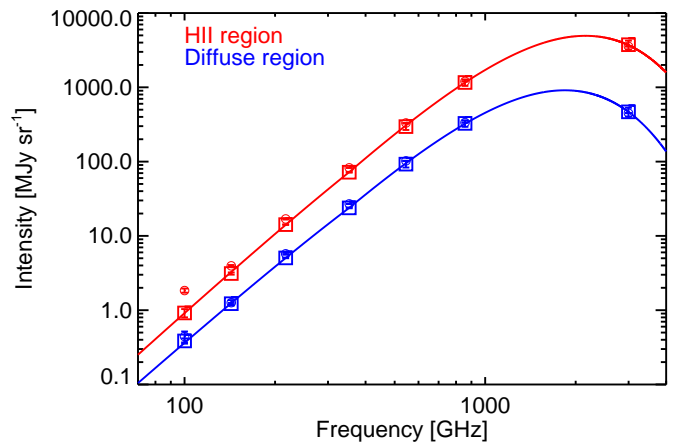}
\caption{Spectra towards the \hii~region complex W42 (red) and a
  diffuse region in the Galactic plane
  centred at $(l{,}b) = (40\pdeg5{,}
0\pdeg0)$ (blue). The circles show the total intensity (corrected for CO
emission) and the squares show the same data after subtraction of the
free-free contribution. All the data points are shown with their
corresponding uncertainties.}
\label{fig:sed1}
\end{figure}

%________________________________________________________________
\section{The dust spectral index from FIR to millimetre wavelengths}
\label{sec:results}

In this section we present the main results of this work, namely the
difference between $\beta_{\rm FIR}$ and $\beta_{\rm mm}$ in the Galactic
plane and how the latter relates to changes in dust temperature and
optical depth. Several tests are performed to assess the robustness of
the results, including a validation of the analysis techniques via
simulations.

\begin{figure*}
\centering
\hspace{-0.8cm}
\includegraphics[scale=0.4]{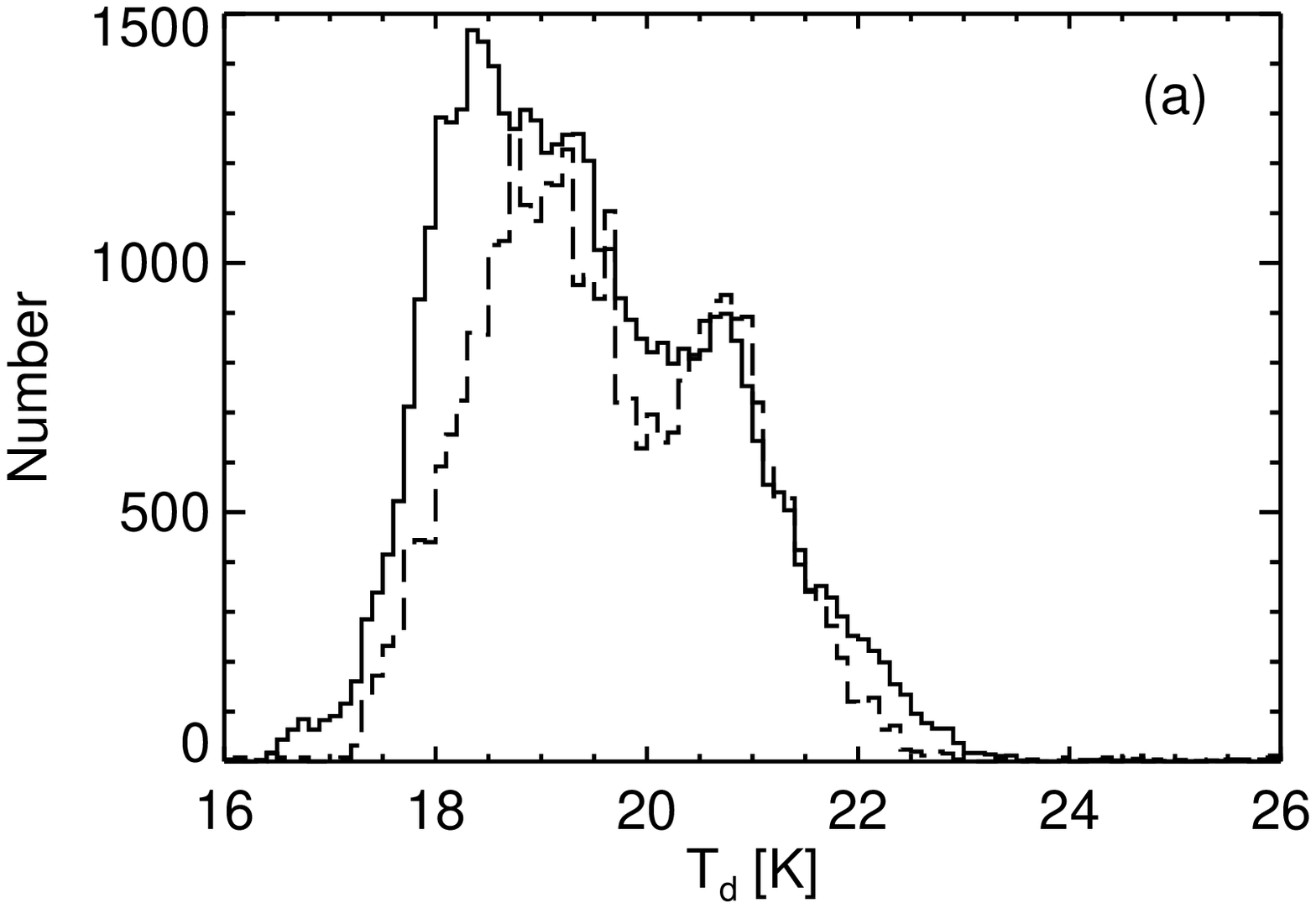}
\includegraphics[scale=0.4]{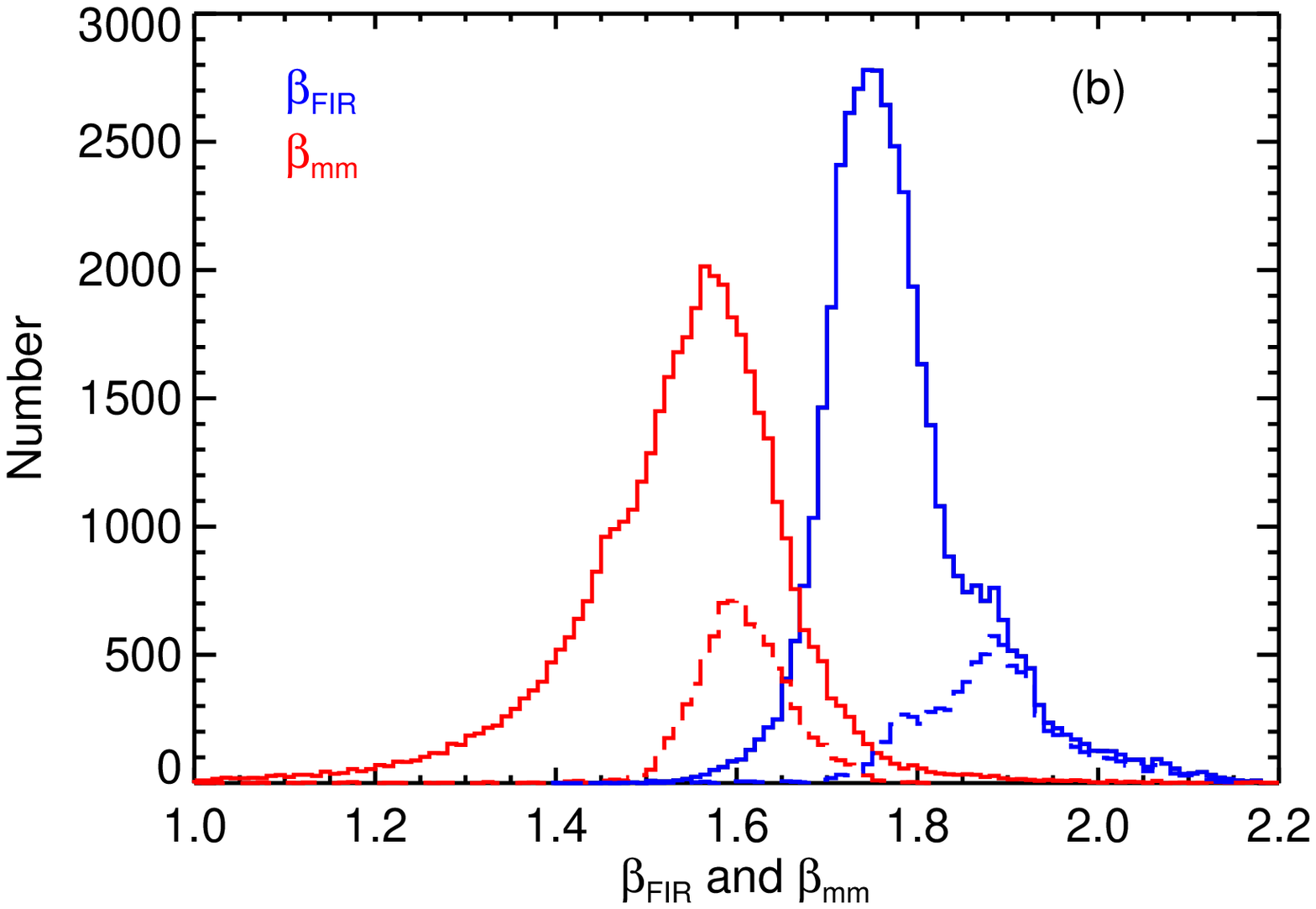}
\vspace{-0.2cm}
\caption{Histograms of the dust temperature (a) and dust opacity
  indices (b) for the $24\degr \times 8\degr$ region. The dashed lines
  in both panels correspond to the pixels where $\tau_{353}
  \geq 4 \times 10^{-4}$. In panel (a) the dashed histogram
    is scaled up by a factor of four. }
\label{fig:hists}
\end{figure*}

\begin{figure*}
\centering
\hspace{-0.8cm}
\includegraphics[scale=0.4]{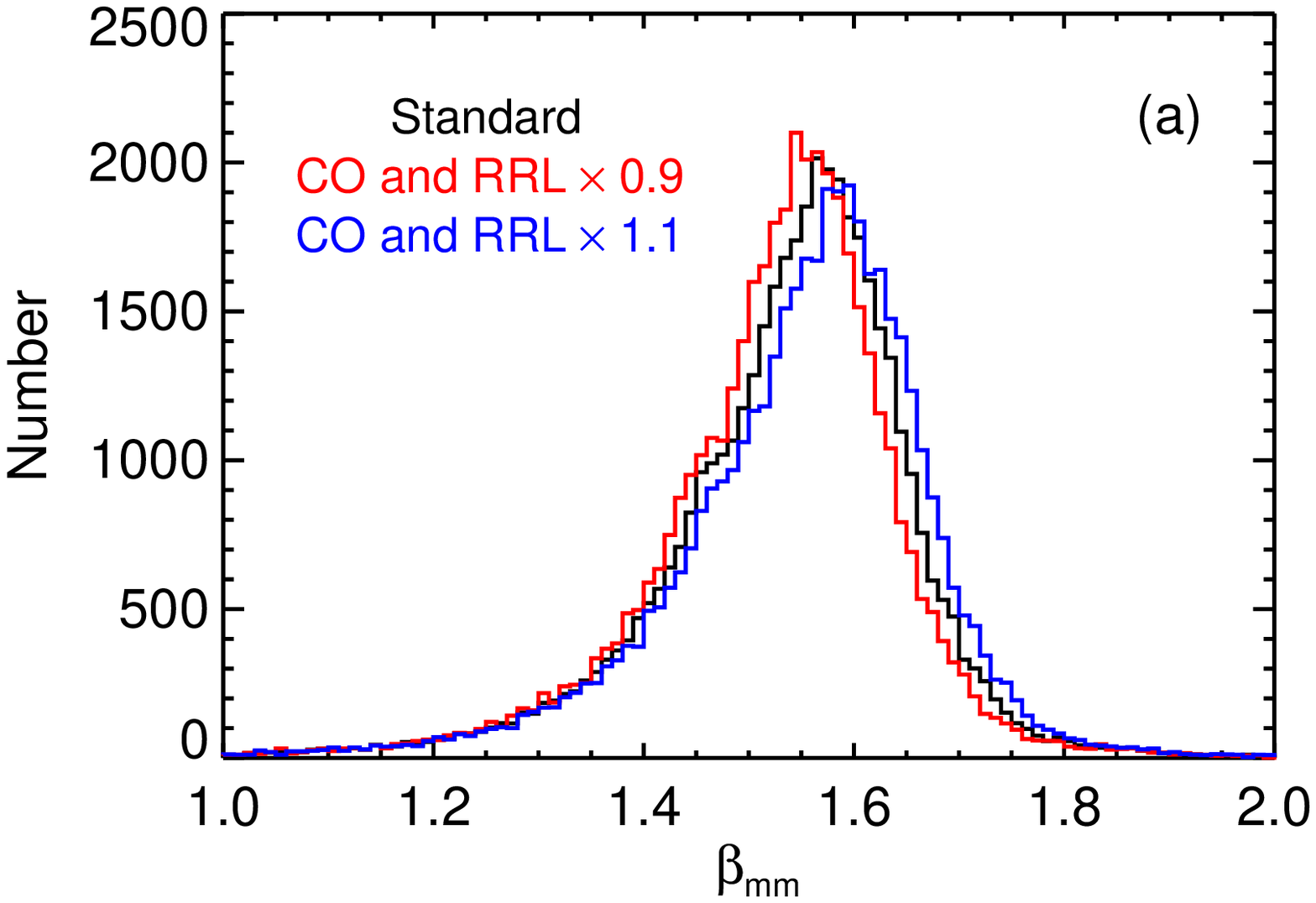}
\includegraphics[scale=0.4]{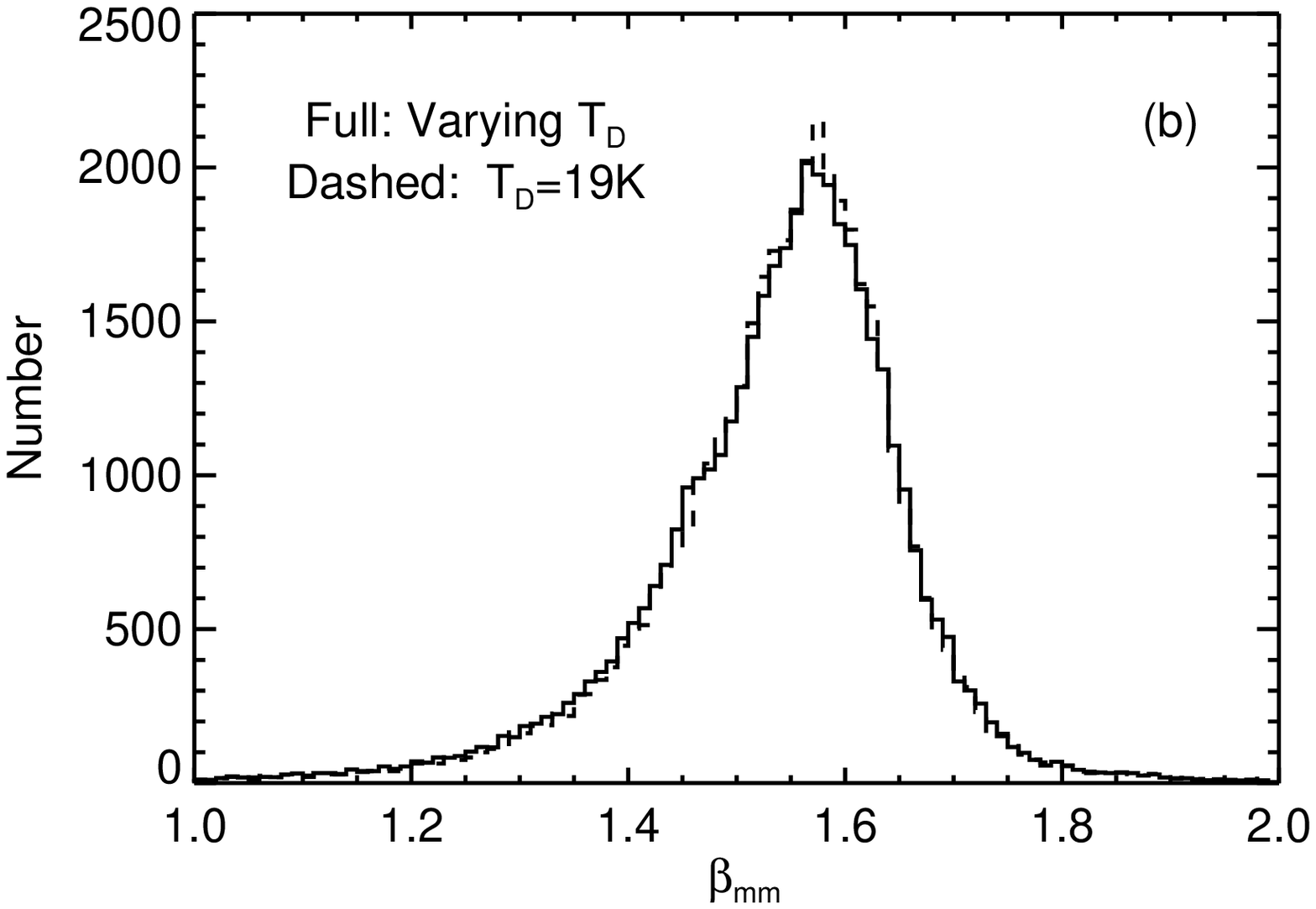}
\vspace{-0.2cm}
\caption{Comparison of the results on $\beta_{\rm mm}$. (a) When the CO and
  free-free corrections vary by 10\,\%. (b) When $T_{\rm d}$ is
  allowed to vary in the fit and also when $T_{\rm d}$ is fixed to a single
  value of 19\,K.}
\label{fig:coffbeta}
\end{figure*}

%________________________________________________________________
\subsection{Flattening of the dust SED}
\label{subsec:res1}

The spectra at the position of a complex of
\hii~regions, G24.5+0.0 (W42), and towards a diffuse region in the
Galactic plane centred at $(l{,}b) = (40\pdeg5{,}0\pdeg0)$ are shown
in Fig. \ref{fig:sed1}. The fitted models are also shown. The effect of subtracting the free-free emission is clearly
visible at 100\,GHz in the spectrum of the \hii~region (compare the
circles with the squares); at
frequencies above 143\,GHz this subtraction is negligible. 
The spectral indices of the \hii~region are $\beta_{\rm FIR}=1.9
 \pm 0.2$ and $\beta_{\rm mm} = 1.7 \pm 0.1$, while for the diffuse region,
 $\beta_{\rm FIR}=1.9 \pm 0.2$ and $\beta_{\rm mm} = 1.6 \pm
 0.1$. These values suggest that the diffuse region has a flatter
 millimetre spectrum than the \hii~region. The uncertainties on the parameters reflect the likely mixture of
 dust components along the line of sight, which have a range of
 temperatures and different properties. The $\chi^{2}$
 values of the fits are 3.1 and 1.1, for the \hii~region and the diffuse
 region respectively, with $N_{\rm dof}=3$\footnote{Degrees of freedom
 (${\rm dof} = N_{\rm points} - N_{\rm parameters}$).}.
%Similarly to the examples in Fig.~\ref{fig:sed1}, 
The $\chi^{2}/N_{\rm dof}$ values across the map are usually lower
than one, meaning that the fits are within the uncertainties of each
point. The uncertainties on the data at frequencies of 217\,GHz and
above are dominated by the calibration uncertainties, 
which are correlated across the channels. 
At these frequencies, the median value of our fit residuals across the map is close to zero and within the overall uncertainties of the data,
 thus indicating that the fits are a good representation of the
 data. At 100 and 143\,GHz the histograms of the percentage
 residual emission are centred at 2\,\% and $-3$\,\%,
 respectively. These values are higher than the 0.5\,\%
 calibration uncertainty at these frequencies but lower than the final
 uncertainties once the noise contribution from
 CMB and the uncertainties associated with free-free and CO templates are included.

The distributions of temperature and spectral
indices fitted for the $24\degr \times 8\degr$ region under study are
shown in Fig.~\ref{fig:hists}. The dust
temperature ranges from 16 to 24\,K, with a median value of 19\,K.
Even though we
are describing the SED with only a single temperature whilst 
a range of temperatures are expected along the line of sight especially in the Galactic plane, the
higher temperature regions found here are associated with \hii~regions, as expected from local heating by their OB stars. Similarly, colder
regions are associated with molecular clouds.

The histograms of $\beta_{\rm FIR}$ and $\beta_{\rm mm}$ are compared in
Fig.~\ref{fig:hists}(b) for the whole $24\degr \times 8\degr$ region. 
The $\beta_{\rm FIR}$ distribution has a median value of 1.76 and a
standard deviation ($\sigma$, corresponding to the 68.3\,\% confidence
interval) of 0.08 and that of $\beta_{\rm mm}$ has a median value of
1.55 with $\sigma = 0.12$. This indicates that the $\beta_{\rm mm}$ distribution is centred at
a lower value and is also broader.
If we select the pixels within $|b| \la 1\degr$, which represents regions with an optical
depth $\tau_{353} \geq 4 \times 10^{-4}$, the corresponding
$\beta_{\rm FIR}$ and $\beta_{\rm mm}$ histograms, shown as dashed lines,
have median values of 1.88, $ \sigma= 0.08$, and
1.60, $\sigma= 0.06$ respectively. The shift in the mean values of both
$\beta_{\rm mm}$ and $\beta_{\rm FIR}$ is related to a variation of
these parameters from the diffuse to the denser medium, as will be
discussed in Sect. \ref{subsec:betamminterp}. The $\beta_{\rm mm}$ values fitted outside the narrow Galactic plane are
affected by CMB fluctuations, which become brighter than the free-free and are not
taken into account in the fit. The impact of the CMB in the $\beta_{\rm mm}$ results will be further
analysed in Sects. \ref{subsec:simulations} and
\ref{subsec:smicacmb}. The dust temperature distribution is
  similar between the $24\degr \times 8\degr$ region and the thin Galactic disk (full
  and dashed lines in Fig. \ref{fig:hists}(a)), with a sharper
  decrease of the latter below 19\,K.

The histogram of $\beta_{\rm FIR}$ in Fig.~\ref{fig:hists} does not
include the effects of the calibration uncertainties, namely its width only
takes into account the variations across
the map. This is an important point when assessing the difference
between $\beta_{\rm FIR}$ and $\beta_{\rm mm}$, as given by Fig.~\ref{fig:hists}.
At frequencies of 353\,GHz and above, where the contribution of CO, free-free and CMB
are negligible compared to dust emission, the data uncertainties are dominated by calibration uncertainties. We performed Monte Carlo
simulations to estimate this effect on $\beta_{\rm FIR}$ and found that,
in 1000 simulations, the dispersion around an input value of 1.75 is
0.17. This value is about twice that measured from the
$\beta_{\rm FIR}$ histograms of Fig.~\ref{fig:hists}. Nevertheless, in the
thin Galactic disk, this does not affect the difference measured
between $\beta_{\rm FIR}$ and $\beta_{\rm mm}$. A further check on the
quality of the SED fits and the importance of including a second
spectral index, $\beta_{\rm mm}$, is given by comparing the
residuals with those resulting from a model with a single $\beta$. When only one spectral index is fit for from 100 to
3000\,GHz, the median value of the residuals across the map is larger
at all frequencies, relative to the two-$\beta$ model. In particular,
the median value of the residuals at 857\,GHz, 11\,\%,
is higher than the calibration uncertainty. We also note that, if
we choose the reference frequency of 545\,GHz, instead of 353\,GHz for
the break in the spectral index, the fits also result in larger residuals
at all frequencies. 

We tested the robustness of the fitted $\beta_{\rm mm}$ against calibration uncertainties in both the CO and the
free-free templates by  varying the subtraction of the
CO and RRL contributions, at all frequencies, by 10\,\% (Sects. \ref{sec:planckdata} and
\ref{subsec:rrldata}). An under-subtraction
of either the CO or the free-free emission could in principle
result in a lower $\beta_{\rm mm}$. However, as Fig.~\ref{fig:coffbeta}(a) illustrates, $\beta_{\rm mm}$ is
essentially insensitive to these variations. This is due to the fact
that dust is the dominant emission component at these frequencies,
combined with the higher uncertainties of the data at 143 and 100\,GHz.

\begin{figure*}
\centering
\vspace{-2cm}
\includegraphics[scale=0.4,angle=90]{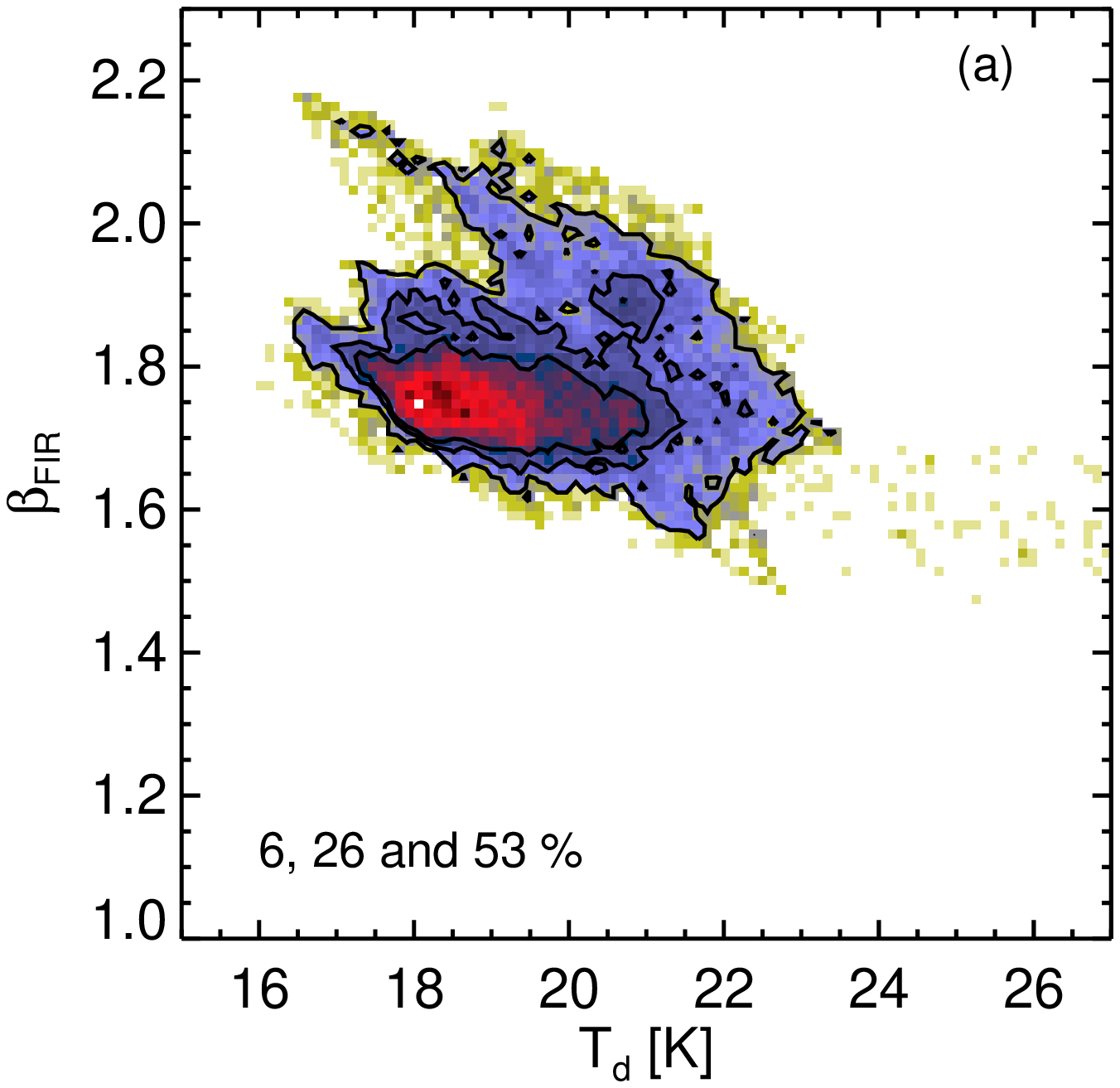}
\hspace{0.3cm}
\includegraphics[scale=0.4,angle=90]{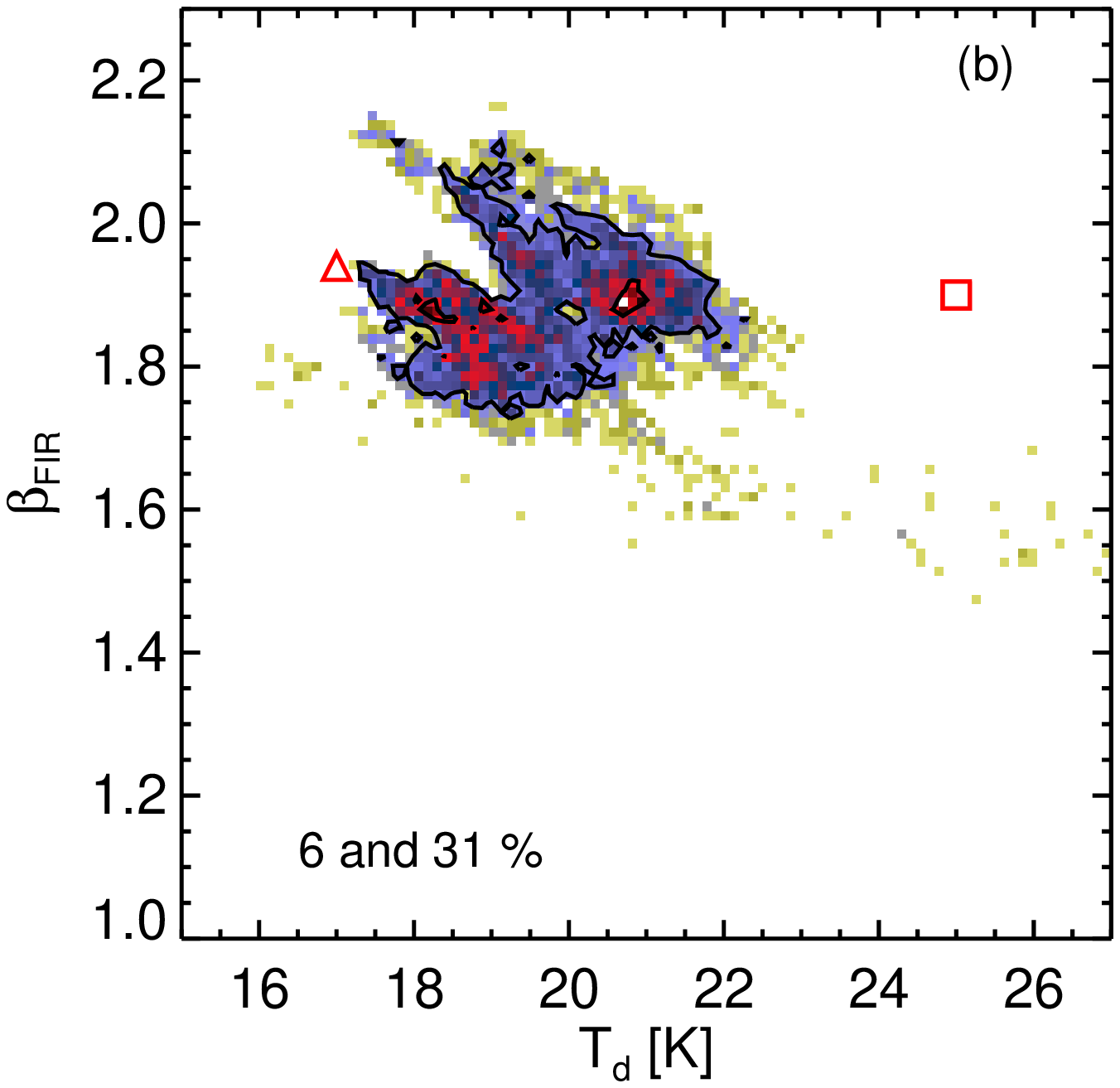}
\hspace{0.3cm}
\includegraphics[scale=0.4,angle=90]{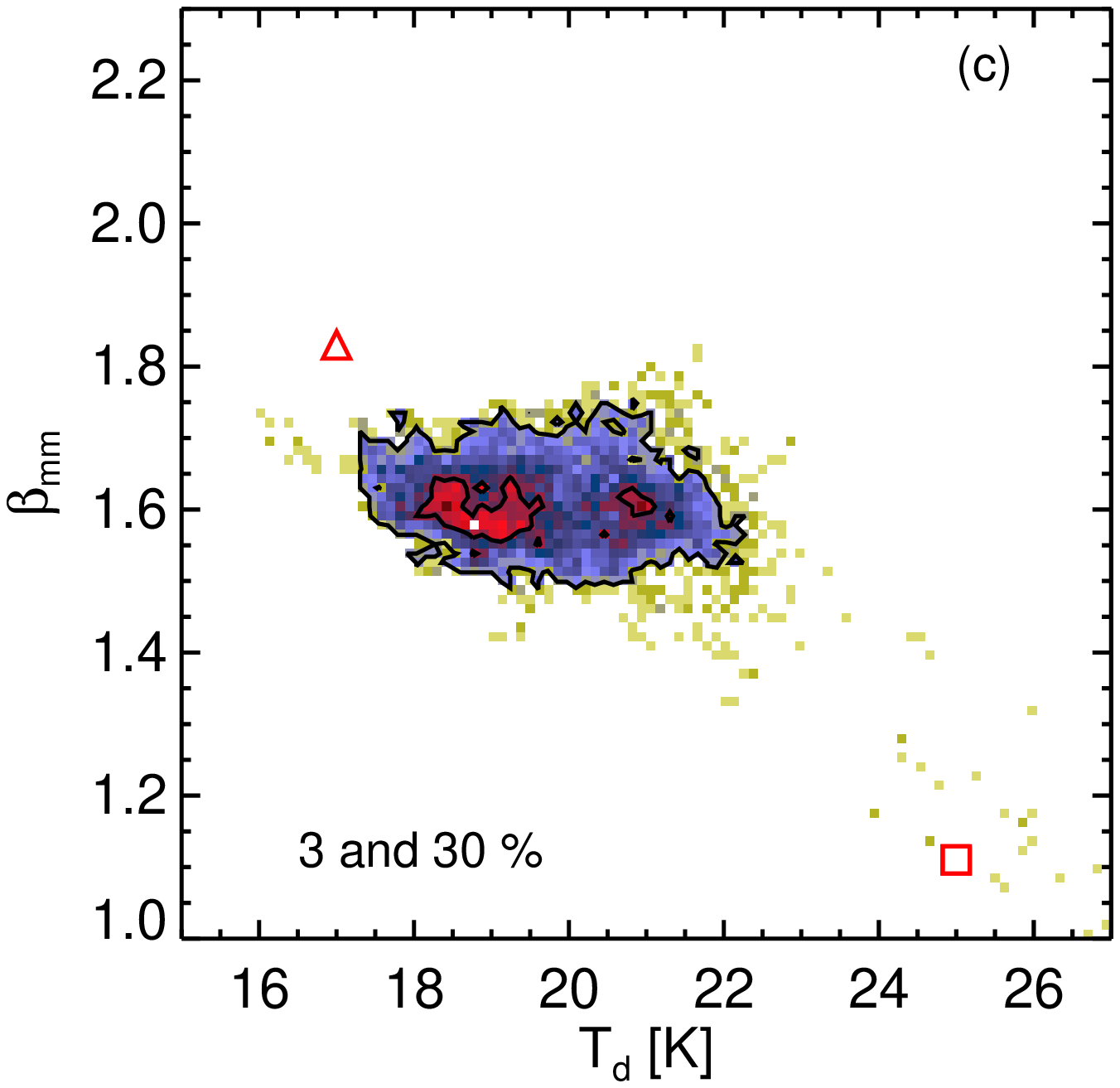}
\caption{Dust spectral indices as a function of temperature. (a)
  $\beta_{\rm FIR}$ versus $T_{\rm d}$ for the whole $24\degr \times 8\degr$ region. (b)
  $\beta_{\rm FIR}$ versus $T_{\rm d}$ for points where $\tau_{353} \geq 4\times
  10^{-4}$. (c) $\beta_{\rm mm}$
  versus $T_{\rm d}$ for points where $\tau_{353} \geq 4\times 10^{-4}$. The
  triangle and square in panels (b) and (c) indicate the values
obtained by fitting the emissivities predicted by
the TLS model \citep{Paradis:2011} for $T_{\rm d}=17$~and 25\,K (see
Sec. \ref{subsec:tls}). The colour scale is logarithmic and it represents the density
  of points. The contours show the densities for the
  cumulated fractions, given by the values in each panel, of the data points, from red to yellow.}
\label{fig:betas-temp}
\end{figure*}

In order to investigate the impact of the dust temperature on
$\beta_{\rm mm}$, we compare the results when $T_{\rm d}$ is fixed to 19\,K
with those when $T_{\rm d}$ is allowed to vary in the SED fit.
These are shown in Fig.~\ref{fig:coffbeta}(b), where it is seen that the distribution of $\beta_{\rm mm}$ is
unaffected when using a constant or varying value of $T_{\rm d}$ across the region.

%________________________________________________________________
\subsection{Variations with temperature and optical depth}
\label{subsec:res3}

An anti-correlation between $\beta$ and $T_{\rm d}$ has been detected in previous observations in a variety of
Galactic regions
\citep{Dupac:2003, Desert:2008, Paradis:2010,planck2011-7.13}. This seems to indicate
that the dust opacity index decreases with temperature, even if part of
this effect can be attributed to data noise and to temperature
mixing along the line of sight
\citep{Sajina:2006,Shetty:2009a,Shetty:2009b,Juvela2012a,
  Juvela2012b}. \citet{Kelly:2012} show that the former can be
mitigated by using a hierarchical Bayesian technique. The distribution
of $\beta_{\rm FIR}$ as a function of
$T_{\rm d}$ is shown in Fig. \ref{fig:beta-fh2}(a), for all the points in the $24\degr \times 8\degr$
region. The Pearson correlation coefficient\footnote{The correlation matrix is
  computed from the covariance matrix of the fit; it measures the intrinsic
  correlation between the uncertainties on the fit parameters.} between the
uncertainties on these two parameters is around $-0.95$ across the
map.
However, such a strong anti-correlation
is not observed in Fig.~\ref{fig:betas-temp}(a), nor in
Fig.~\ref{fig:betas-temp}(b), where $\beta_{\rm FIR}-T_{\rm d}$ is
plotted for the thin Galactic disk, $|b| \la 1\degr$. Therefore,
there is a real variation of $\beta_{\rm FIR}$ across this region which
decreases the anti-correlation trend generated by the data noise.

The distribution of $\beta_{\rm mm}$ as a function of $T_{\rm
  d}$, for the thin Galactic disk, where $\tau_{353} \geq 4 \times
10^{-4}$, is shown in Fig. \ref{fig:betas-temp}(c). For this region, the correlation coefficient between the
uncertainties on $\beta_{\rm mm}$ and $T_{\rm d}$ varies between
$-0.06$ and $-0.03$.
A value of $-0.5$ is reached outside the Galactic disk,
where the signal-to-noise ratio decreases due to the CMB noise term
included in the data uncertainties. Thus, Fig. \ref{fig:betas-temp}(c)
indicates that there is no evident trend of $\beta_{\rm mm}$ with
$T_{\rm d}$ (as it is also seen by comparing the corresponding maps in
Fig. ~\ref{fig:betamm_map}). We note that the range of temperatures that we are probing is
limited, about 6\,K, which may be due to temperature mixing along the
line of sight and local temperature increases around the heating sources present in the Galactic plane.

\begin{figure}
\centering
\vspace{-2.5cm}
\includegraphics[scale=0.55,angle=90]{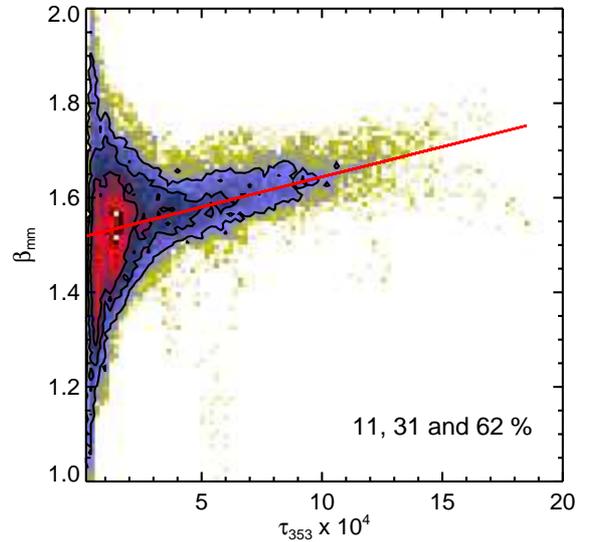}
%\vspace{-1cm}
\caption{The distribution of $\beta_{\rm mm}$ as a function of $\tau_{353}$, for the whole
  region. The red line gives
  the best linear fit for $\tau_{353} \geq 4\times 10^{-4}$ (see
  text). The colour scale is logarithmic and it represents the density
  of points. The three contours show the densities for a
  cumulated fraction of 11, 31, and 62\,\%  of the data points, from red to yellow.}
\label{fig:beta-tau}
\end{figure}

\begin{figure*}
\centering
\vspace{-2cm}
\includegraphics[scale=0.4,angle=90]{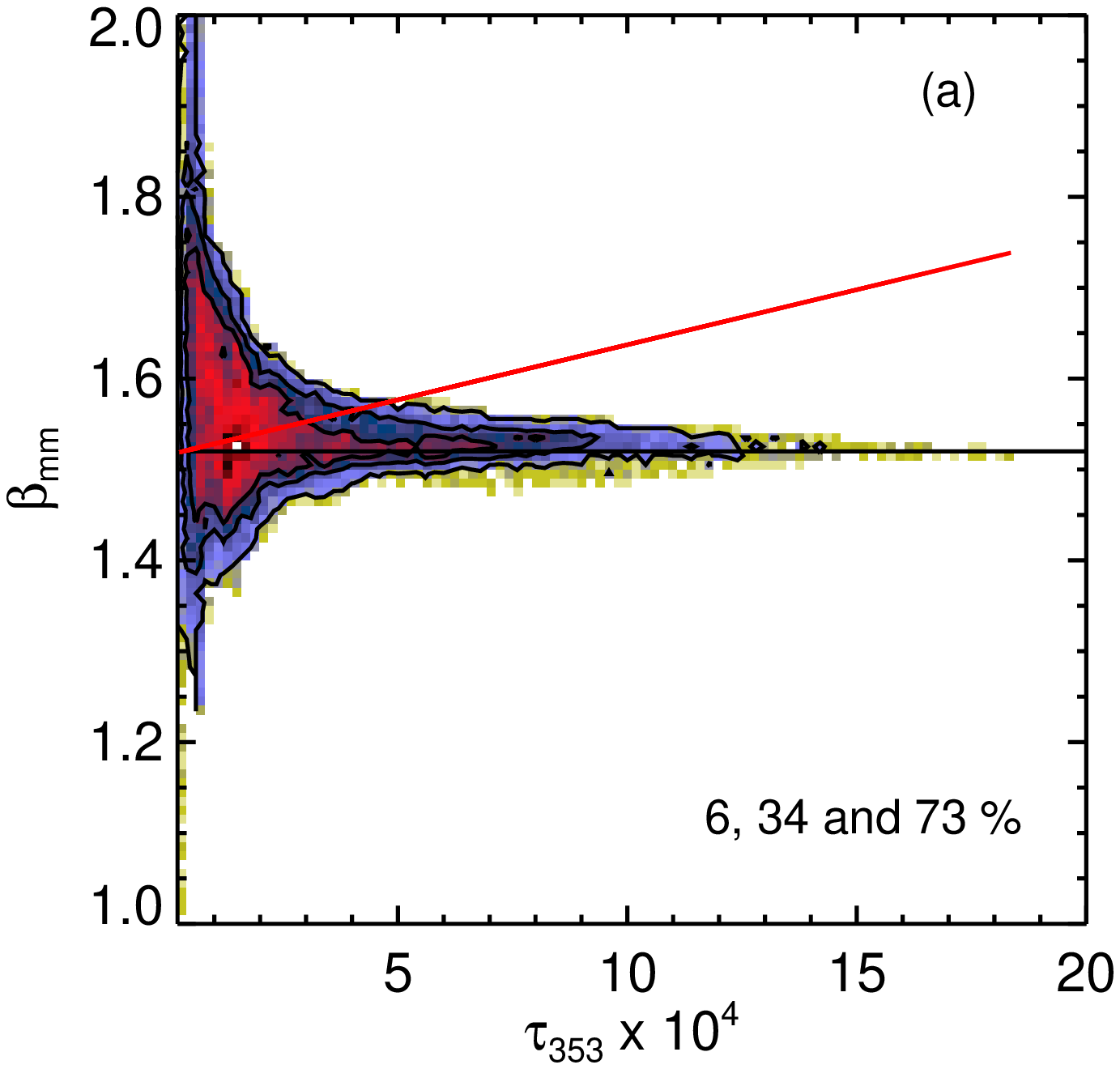}
\hspace{0.5cm}
\includegraphics[scale=0.4,angle=90]{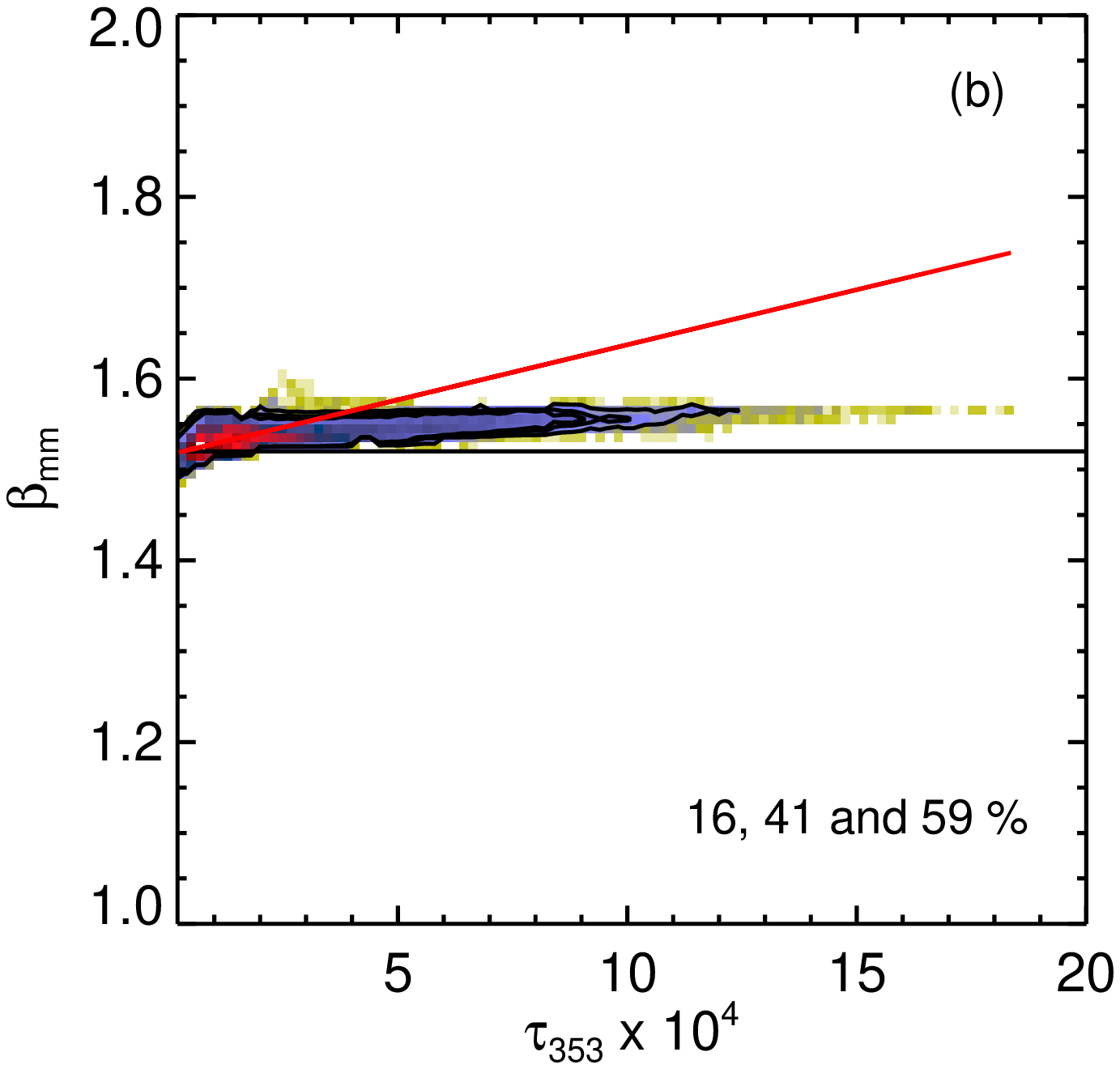}
\caption{The dust opacity index $\beta_{\rm mm}$ recovered from the
  simulated maps, as a function of the input $\tau_{353}$. The simulated maps in (a) have dust
  emission and CMB, and in (b) consist of dust, free-free and CO
  emission. The black line shows the input $\beta_{\rm mm}$ of 1.52 in
  each case; the red lines give the $\beta_{\rm mm}$--$\tau_{353}$ relationship
  derived in Sect.~\ref{sec:results}. The colour scale is logarithmic and it represents the density
  of points. The three contours show the densities for the
  cumulated fractions, given by the values in each panel, of the data points, from red to yellow.}
\label{fig:simul}
\end{figure*}

We find a correlation between $\beta_{\rm mm}$ and the optical depth
$\tau_{353}$, which is shown in Fig.~\ref{fig:beta-tau}. We
  note that $\beta_{\rm mm}$ is an intrinsic parameter related to the physics of dust while
  $\tau_{353}$ scales with the column density of interstellar matter.
In Sect.~\ref{sec:interpretation} we will describe this empirical
correlation in terms of the type of matter rather than the quantity of
matter along the line of sight, given by $\tau_{353}$.
The scatter in $\beta_{\rm mm}$ at low optical depth values, or
$|b| \ga 1\degr$, is due to the CMB, as discussed above. For $\tau_{353} \geq 4\times 10^{-4}$, 
$\beta_{\rm mm}$ increases in the highest
optical depth regions, as foreshadowed by the results of
Fig.~\ref{fig:sed1}, which showed an increase of $\beta_{\rm mm}$ from
the diffuse to the \hii~region. A linear fit to the data, for $\tau_{353} \geq 4\times 10^{-4}$,
 gives $\beta_{\rm mm} = (1.52 \pm 0.01) + (128 \pm 2)\times
 \tau_{353}$, where we have used the {\tt IDL} routine {\tt regress} to perform the linear
 regression fit, including only the errors
on $\beta_{\rm mm}$. We note that the errors on $\tau_{353}$ are much
lower than those on $\beta_{\rm mm}$. Moreover, the Pearson correlation
coefficient between the uncertainties on these two
parameters for $\tau_{353} \geq 4\times 10^{-4}$, varies between 0.03 and 0.08. This indicates that it is
unlikely that the correlation observed is due to data noise.

The uncertainty on $\beta_{\rm mm}$ is statistical; including the
systematic uncertainties introduced by the CMB, CO and free-free
components, which are presented in Sect.~\ref{subsec:simulations}, the correlation is
\begin{equation}
\beta_{\rm mm} = (1.52 \pm 0.02) + (128 \pm 24)\times \tau_{353}.
\label{eq:betamm-tau}
\end{equation}

%________________________________________________________________
\subsection{Validation with simulations}
\label{subsec:simulations}

In order to test the robustness of our fitting procedure against
possible biases on $\beta_{\rm mm}$ associated with the separation of
dust emission from CMB, free-free and CO, we apply our routine to
simulated maps.

The first simulated maps include dust emission
and CMB. We fix $T_{\rm d}$ to 19\,K, $\beta_{\rm FIR}$ to 1.75 and
$\beta_{\rm mm}$ to 1.52 across the region. The distribution of $\tau_{353}$ is that obtained from the fit to the data. We reproduce
the dust maps at each frequency with a modified blackbody
law and add them to the CMB map, reproduced from the best-fit
$\Lambda$CDM model. We then apply the SED fitting routine and recover $\beta_{\rm mm}$
as a function of the input optical depth as shown in
Fig.~\ref{fig:simul}(a).  A linear fit to the points
with $\tau_{353} \geq 4 \times
10^{-4}$ gives $\beta_{\rm mm}=1.53 - 8 \times \tau_{353}$. As discussed in Sect.~\ref{sec:results},
the scatter on the $\beta_{\rm mm}$ values for $\tau_{353} \la 4 \times
10^{-4}$ is created by the CMB when this component is not taken into account in the
fit. Moreover, even though it is a minor contributor in the Galactic
disk, the CMB also affects the results for higher values of $\tau_{353}$, broadening the $\beta_{\rm mm}$
distribution around the input value of 1.52 by 0.01. Thus this result
shows that if there is no intrinsic correlation between these two parameters only a limited
correlation will be detected. More importantly, Fig.~\ref{fig:simul}(a) shows that the CMB is not responsible for
the $\beta_{\rm mm}$--$\tau_{353}$ correlation derived in the previous
section.

We also tested our results for a possible bias introduced by incorrect
subtraction of CO and free-free emission from simulated dust maps
produced as described above.
For that we use the MILCA CO
maps (Sect.~\ref{sec:planckdata}) and subtract 10\,\% of their
emission at 100, 217, and 353\,GHz. Similarly we remove 10\,\% of the
RRL free-free emission from the simulated dust
maps at all frequencies. Such a correction steepens the dust spectrum, as we can see
from the results of Fig.~\ref{fig:simul}(b). A linear fit to the
points gives $\beta_{\rm mm}=1.53 + 22 \times \tau_{353}$.
This is not, however,
capable of reproducing the much steeper slope of $\beta_{\rm mm}$ with the
dust optical depth. For that to be the case, both the CO and RRL maps
would have to be systematically underestimated by 30\,\%. 

\begin{figure}
\hspace{-0.4cm}\vspace{0.1cm}
\includegraphics[scale=0.36,angle=90]{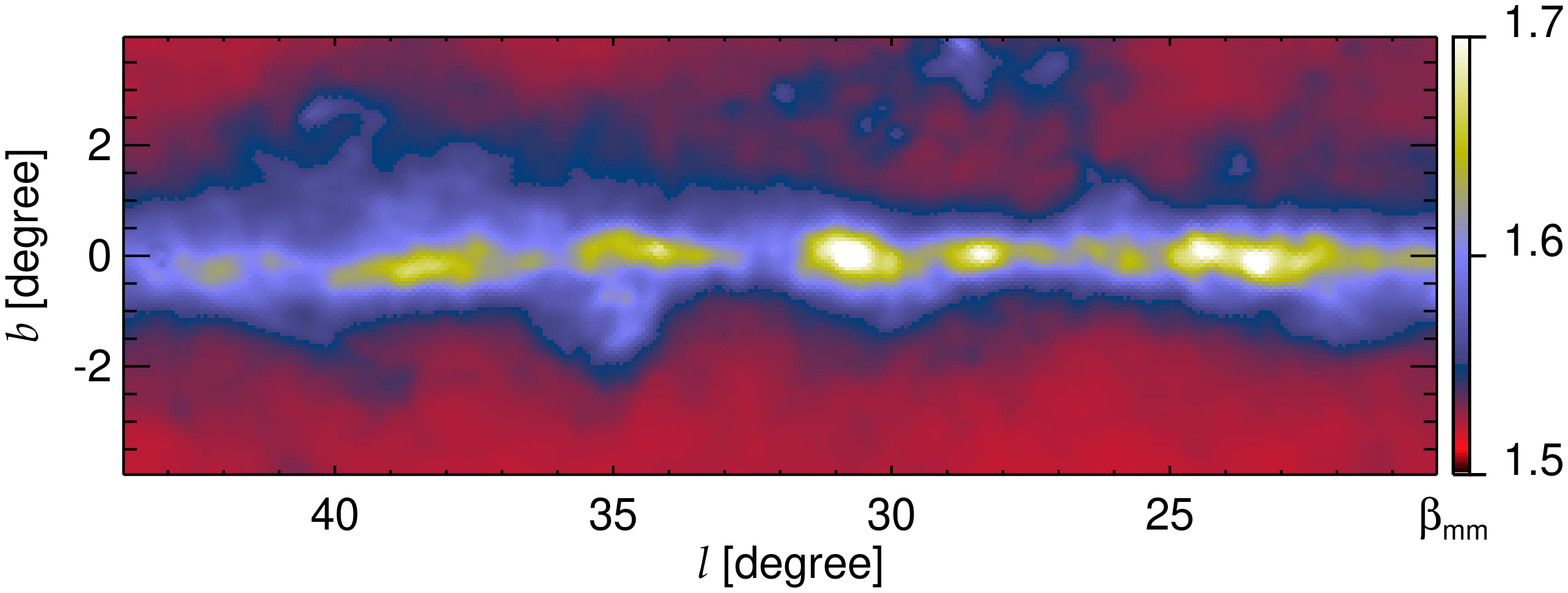}
\vspace{0.1cm}\hspace*{-0.4cm}
\includegraphics[scale=0.36,angle=90]{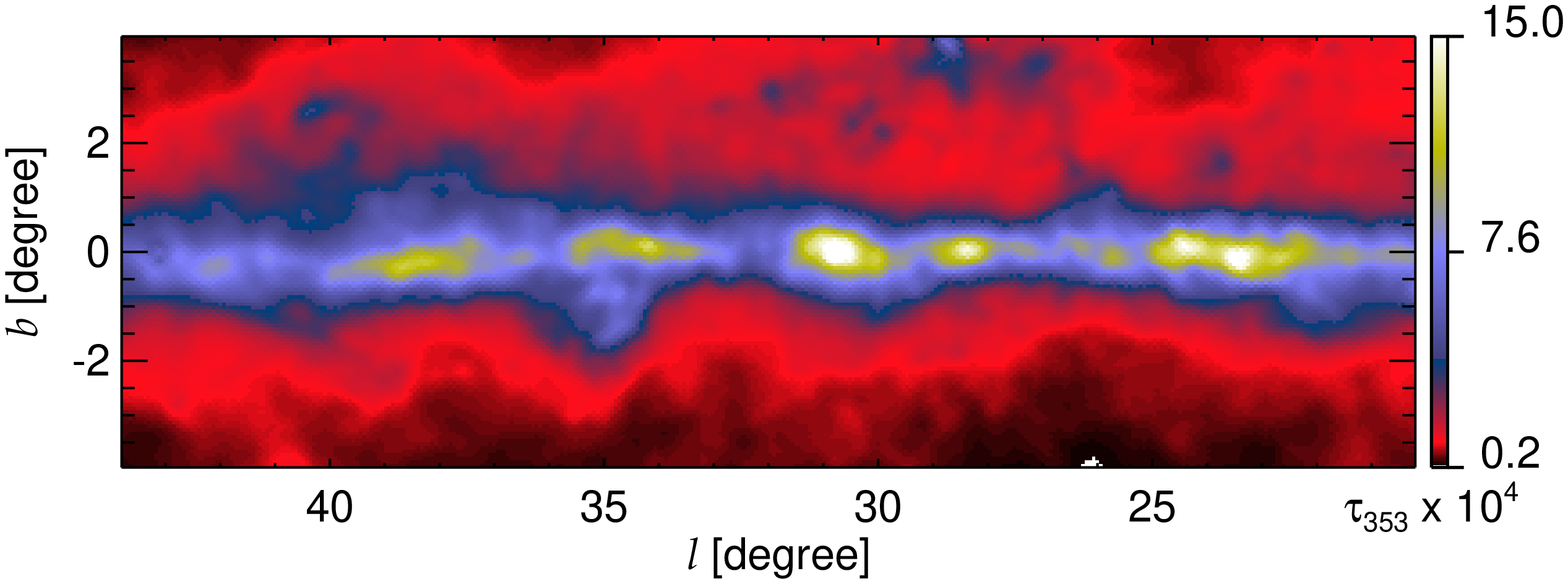}
\hspace*{-0.4cm}
%\hspace*{4.5cm}
\includegraphics[scale=0.36,angle=90]{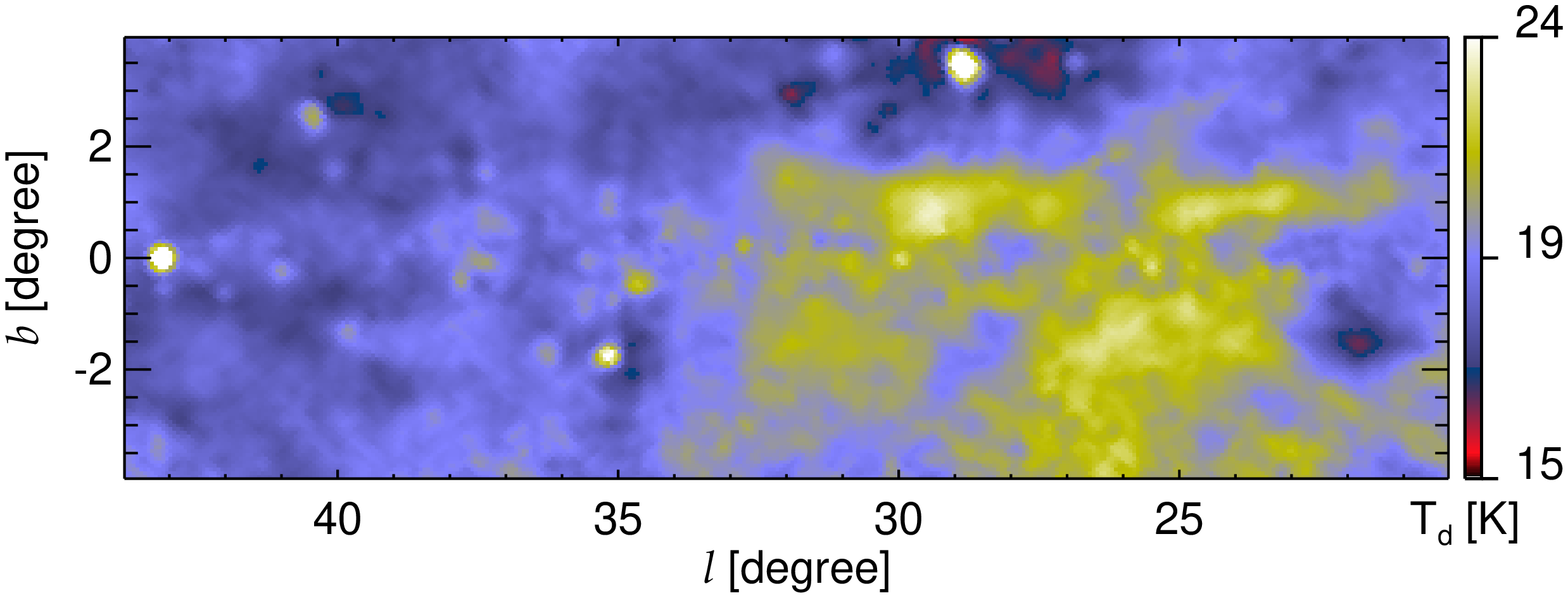}
\vspace{0.1cm}
\caption{Maps of the dust parameters. \emph{Top}: $\beta_{\rm mm}$, which
  results from the best linear fit to the correlation with optical
depth given by Eq. (\ref{eq:betamm-tau}); \emph{Middle}: $\tau_{353}$;
\emph{Bottom}: $T_{\rm d}$.}
\label{fig:betamm_map}
\end{figure}

We thus conclude that neither the uncertainty in the CO and free-free
correction of the maps nor omitting the CMB in the spectral fits is
responsible for the correlation of $\beta_{\rm mm}$ with
$\tau_{353}$. 
Finally, Fig.~\ref{fig:betamm_map} shows the map of $\beta_{\rm mm}$, estimated using Eq.
(\ref{eq:betamm-tau}), which presents the same structure as the map of
$\tau_{353}$. On the other hand, there is no apparent correlation
between the maps of $\beta_{\rm mm}$ and $T_{\rm d}$, as it will be discussed in
Sect. \ref{sec:interpretation}.

\begin{figure}
\centering
\vspace{-2cm}
\includegraphics[scale=0.55,angle=90]{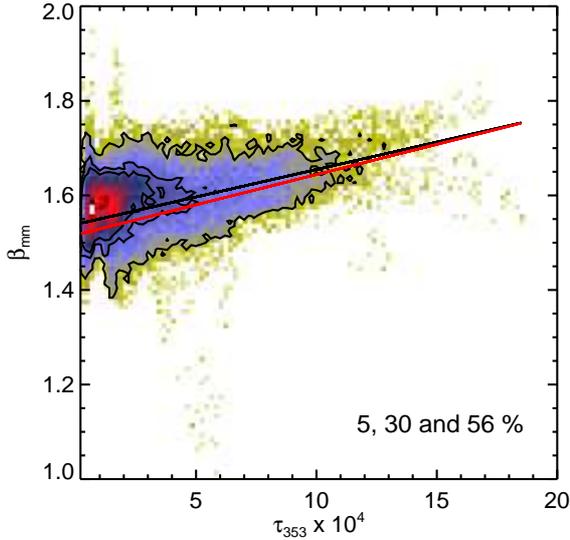}
\caption{The distribution of $\beta_{\rm mm}$ as a function of $\tau_{353}$, for the whole
  region, when the {\it SMICA} CMB map is subtracted from each channel
  map. The black line gives the $\beta_{\rm mm}$--$\tau_{353}$ relationship
  derived from a fit to the points where $\tau_{353} \geq 4 \times
10^{-4}$, compared to that estimated in Sect.~\ref{sec:results} and
given by the red line. The colour scale is logarithmic and it represents the density
  of points. The three contours show the densities for a
  cumulated fraction of 5, 30, and 56\,\% of the data points, from red to yellow.}
\label{fig:smicacmb}
\end{figure}

%________________________________________________________________
\subsection{Using the \Planck~{\it SMICA} CMB map}
\label{subsec:smicacmb}

In this section we compare our results with those obtained when the
{\it SMICA} CMB map is
subtracted from each channel map before fitting the dust spectra with a
modified blackbody. The resulting $\beta_{\rm mm}$ as a function of
$\tau_{353}$ is shown in Fig.~\ref{fig:smicacmb}. As expected, the scatter on
$\beta_{\rm mm}$ at low values of $\tau_{353}$ decreases, due to the
subtraction of the CMB from the total emission. A linear fit to the points
with $\tau_{353} \geq 4 \times
10^{-4}$ gives $\beta_{\rm mm}=(1.54\pm0.01) + (116\pm1) \times
\tau_{353}$, consistent with Eq. (\ref{eq:betamm-tau}). This confirms that the CMB
fluctuations, here measured with the {\it SMICA}
solution, are indeed a small contribution in the Galactic disk and do not affect the main results of the
work.

%________________________________________________________________
%________________________________________________________________
\section{Towards a physical interpretation of the millimetre dust emission}
\label{sec:interpretation}

In this section we compare our results with
predictions from current dust models and interpret the empirical
relation found between $\beta_{\rm mm}$ and $\tau_{353}$.

%________________________________________________________________
\subsection{Dust models}
\label{subsec:models}

%________________________________________________________________
\subsubsection{Silicate-carbon models}
\label{subsec:dl07}

We start by comparing our results with the predictions of two commonly
used dust models, DL07 \citep{DraineLi:2007} and DustEM 
\citep{Compiegne:2011}. In particular, we want to investigate whether such models, with
two populations of grains dominating the emission at long wavelengths,
can explain the flattening of the dust spectrum detected in the
present work. Both models use the same optical properties for
silicates, for which the opacity scales as $\nu^{1.6}$, for $\lambda
\ga 250$\microns~or $\nu \la 1200$\,GHz. For the carbon grains DL07 uses the
optical properties of graphite, with a spectral index of 2, whereas DustEM uses the laboratory measurements of amorphous carbon, for
which the spectral index is 1.6. 
We use both models to predict the emission in the photometric bands
considered in this work, namely \textit{IRAS} 100\micron~and HFI, taking the
standard size distribution for the diffuse Galactic emission. In order to
reproduce conditions closer to those in the Galactic plane, we generate the SEDs for $G_{0}$ values of 1, 2 and 4, where $G_{0}$ is the scaling applied to the standard ISRF of \citet{Mathis:1983}. We then fit the spectra in the same way as the data, namely with a modified blackbody law and two spectral
indices, $\beta_{\rm FIR}$ and $\beta_{\rm mm}$. The results are shown in
Table \ref{table:dl07}. First we
note that when the radiation field is higher, the peak of the SED is moved to higher frequencies, where the opacity
spectral index of silicate grains is larger than 1.6. This can explain the slight increase in $\beta_{\rm FIR}$ with $G_{0}$.
The results also show
that $\beta_{\rm mm}$ is lower than $\beta_{\rm FIR}$, for both
models. Moreover, $\beta_{\rm mm}$ does not seem to
vary with $G_{0}$ or with $T_{\rm d}$. 
We find that such models, including two dust
components with different opacities can in principle explain the flattening
of the dust emission, even if the $\beta$ values recovered from their spectra are lower
than those measured from the data.

\begin{table}[tmb]
\begingroup
\newdimen\tblskip \tblskip=5pt
\caption{Results from a modified blackbody fit to the DL07 and DustEM
  spectra for different scalings of the ISRF, given by $G_{0}$.}
\label{table:dl07}
\nointerlineskip
\vskip -3mm
\footnotesize
\def\leaderfi1{\leaders\hbox to 5pt{\hss.\hss}\hfil}
\setbox\tablebox=\vbox{
   \newdimen\digitwidth 
   \setbox0=\hbox{\rm 0} 
   \digitwidth=\wd0 
   \catcode`*=\active 
   \def*{\kern\digitwidth}
   \newdimen\signwidth 
   \setbox0=\hbox{+} 
   \signwidth=\wd0 
   \catcode`!=\active 
   \def!{\kern\signwidth}
   \halign{\hbox to 0.7in{#\leaderfil}\tabskip 8pt&
     \hfil#\hfil\tabskip 8pt&
     \hfil#\hfil\tabskip 8pt&
     \hfil#\hfil\tabskip 16pt&
     \hfil#\hfil\tabskip 8pt&
     \hfil#\hfil\tabskip 8pt&
     \hfil#\hfil\tabskip 0pt\cr
     \noalign{\doubleline\vskip1pt}
\omit & \multispan3\hfil DL07 \hfil &  \multispan3\hfil DustEM \hfil \cr 
\noalign{\vskip -3pt} 
\omit&\multispan3\hrulefill&\multispan3\hrulefill\cr
 \omit\hfil$G_{0}$\hfil& $T_{\rm d}$ [K]& $\beta_{\rm FIR}$& $\beta_{\rm mm}$&  $T_{\rm d}$ [K]& $\beta_{\rm FIR}$& $\beta_{\rm mm}$\hfil\cr
\noalign{\vskip 3pt\hrule\vskip 5pt}
1& 21.8& 1.65& 1.43& 20.1& 1.58& 1.48\cr
2& 24.1& 1.70& 1.43& 22.4& 1.62& 1.47\cr
4& 26.7& 1.73& 1.43& 25.0& 1.64& 1.46\cr
\noalign{\vskip 5pt\hrule\vskip 3pt}}}
\endPlancktable
\endgroup
\end{table}

%________________________________________________________________
\subsubsection{Two-Level System}
\label{subsec:tls}

The Two-level System (TLS, \citealt{Meny:2007}) model has been proposed to explain the flattening of the dust
emission and its evolution with temperature. This
model consists of three mechanisms which
describe the interaction of electromagnetic waves with an amorphous solid. These are temperature-dependent and
important in the sub-millimetre, for the range of
temperatures relevant to this work. \citet{Paradis:2011}
use the TLS model to fit the spectrum of the diffuse Galactic emission
as well as the spectra of the Archeops sources
\citep{Desert:2008}. Within this model, the opacity spectral index
decreases with increasing temperature. We compare the emissivities
predicted by TLS and given in \citet{Paradis:2011} with our results for the relevant
photometric bands. In particular, we
select two spectra, with $T_{\rm d}$ of 17 and 25\,K, within the range
of temperatures probed in the present work. We apply our fitting
routine to the TLS SEDs to recover $\beta_{\rm FIR}$ and
$\beta_{\rm mm}$, which are shown in Figs. \ref{fig:betas-temp}(b) and (c).
The resulting $\beta_{\rm FIR}$ values are within the range found in this
work, showing a small variation with temperature. However, that is not
the case for $\beta_{\rm mm}$. The values predicted by the TLS model are
not within the range of values found in the Galactic disk, and show a
steep dependence with temperature. We note that the TLS emissivities
used here were computed for a given set of parameters, derived from
the combined fit of the diffuse medium emission and the SEDs of the Archeops compact sources (third set of
parameters in Table 4 of \citealt{Paradis:2011}). Moreover, they are
derived for a single grain, rather than for a grain size
distribution. 
We could argue that the dust temperature estimated
from the modified blackbody fit used here
is not comparable with that derived from the TLS model. However
\citet{Paradis:2012} show that both temperatures agree up to about 25\,K. Still, we note that $T_{\rm d}$ obtained
from their modified blackbody fit assumes $\beta_{\rm FIR}=2$.
We conclude that the TLS model predicts variations of $\beta_{\rm mm}$
which are not apparent in the data. We note, however, that the range
of temperatures sampled by the data is limited and that if spatial
variations of the TLS amplitude, related to the amorphous structure of
the grains, were allowed they could easily hide the temperature
dependence of $\beta_{\rm mm}$ in the data.

\begin{figure}
\centering
\includegraphics[width=1\columnwidth]{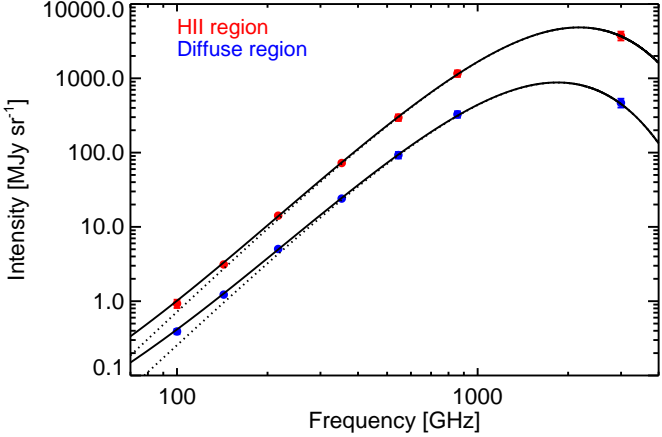}
\caption{Spectra towards the \hii~region complex W42 (red) and a
  diffuse region in the Galactic plane centred at
  $(l{,}b)=(40\pdeg5{,}0\pdeg0)$ (blue). The circles show the total
  intensity, with their corresponding uncertainties. The dotted lines
represent the modified blackbody model, where one single opacity
spectral index is fitted to \textit{IRAS} 100\microns\ and HFI 857, 545,
and 353\,GHz data. The solid lines represent the total emission,
including the contribution from metallic
dust particles \citep{DraineHensley:2013}, at the same temperature
$T_{\rm d}$. This contribution to the total emission
at 100\,GHz is 41\,\% and 63\,\% for the \hii~and diffuse regions, respectively.}
\label{fig:drainemodel}
\end{figure}

%________________________________________________________________
\subsubsection{Magnetic Dipole Emission}
\label{subsec:mdp}

The dust emission of the Small Magellanic Cloud (SMC) shows
a pronounced flattening towards millimetre wavelengths
\citep{Israel:2010,Bot:2010,planck2011-6.4b}, which, as proposed by
\citet{DraineHensley:2012b}, can be explained by magnetic dipole emission from
metallic particles. In this section we test the magnetic dipole
emission model with the Galactic plane data, which also show excess
emission at millimetre wavelengths, even if not as substantial as that
observed in the SMC. According to the model of
\citet{DraineHensley:2012b}, the iron missing from the gas phase can be locked up in solid grains,
either as inclusions in larger grains, in which case they are at the same temperature as the other dust in the diffuse interstellar
  medium (ISM),  $T_{\rm d} \approx 18$\,K, or as free-flying
nanoparticles, which then have a higher temperature, $T_{\rm d} \approx 40$\,K. The
  emission spectrum of these particles above a resonance frequency, $\nu \sim
15$\,GHz, and below 353\,GHz, is close to that of a blackbody. In order to test this model, we fit the dust SEDs in the region under study with
a modified blackbody of a single opacity index $\beta_{\rm FIR}$. Its
value is determined using the \textit{IRAS} and HFI 857, 545, and 353\,GHz points
and then used to extrapolate the emission to lower frequencies. We include a blackbody spectrum, at the same temperature $T_{\rm d}$, to
represent the metallic particles as inclusions in larger grains, which
will account for the excess emission. We find
that, at 100\,GHz, the ratio between the emission from the iron dust
particles and that from the modified blackbody, $r_{100}$, has a
median of 63\,\% across the thin Galactic disk, with a standard deviation
of 24\,\%. The spectra of the
same \hii~and diffuse regions as in Fig.~\ref{fig:sed1} are shown in
Fig. \ref{fig:drainemodel},  for which
$r_{100}$ is $(41\pm8)$\,\% and $(63\pm7)$\,\%, respectively. The contribution by the metallic particles is higher for
the diffuse region since its SED is flatter at lower frequencies than
that of the \hii~region (Sect.~\ref{sec:results}). The fraction obtained here is within the range of plausible values for
  magnetic dipole emission within the model of
  \citet{DraineHensley:2013}, and smaller than that fitted for the SMC
  \citep{DraineHensley:2012b}. In a similar analysis performed at
  high Galactic latitudes, \citet{planck2013-XVII} find
  $r_{100}=26\pm6$\,\%. The lower ratio follows the lower difference
  between their mean values for the FIR and millimetre spectral
  indices, $\beta_{\rm FIR}=1.65$ and $\beta_{\rm mm}=1.53$.

%________________________________________________________________
\subsection{Correlation between $\beta_{\rm mm}$ and $\tau_{353}$}
\label{subsec:betamminterp}

In this section we attempt to provide a phenomenological interpretation of the
empirical correlation detected between $\beta_{\rm mm}$ and
$\tau_{353}$. As mentioned in Sect.~\ref{subsec:res3}, the dust optical
depth provides a measure of the quantity of matter along the
line of sight, which may be atomic or molecular and which has the
contribution of both dense and diffuse media. We suggest that
this variation of $\beta_{\rm mm}$ with the dust optical depth can be translated into an
evolution with the fraction of molecular gas along the
line of sight. 

The fraction of molecular gas is given by $f_{\rm H_{2}}=2 
N_{\rm H_{2}}/N_{\rm H}^{\rm tot}$. The column density of molecular hydrogen
can be estimated by using the conversion factor $X_{\rm CO} =   {
N}_{\rm H_{2}}/{ I}_{\rm CO}$, where ${ I}_{\rm CO}$ is the $^{12}$CO $J$=1$\rightarrow$0 integrated
line intensity. The Galactic $X_{\rm CO}$ conversion factor has been
estimated in a variety of ways, including the use of optically thin
tracers of column density such as dust emission, molecular and atomic
lines, as well as using $\gamma$-ray emission. \citet{Bolatto:2013}
give $X_{\rm CO}=2.0\times10^{20}$\,cm$^{-2}$\,(K\,\kms)$^{-1}$,
with 30\,\% \ uncertainty, as the recommended
value to use in Galactic studies. We can obtain an estimate of $X_{\rm CO}$ with the present data using the dust optical depth and
the CO emission provided by the MILCA map. For that we need to
include the dust specific opacity, or absorption cross-section per
unit gas mass,
of the molecular gas after removing the contribution of the
atomic gas to the dust optical depth.
Dust properties are known to evolve from the diffuse ISM to
the higher density environment of molecular clouds, giving rise to an
enhancement of the dust specific opacity \citep{planck2011-7.12,planck2011-7.13,planck2013-p06b}. One possible
explanation is grain coagulation
\citep{Stepnik:2003,Kohler:2012}. The dust specific opacity appears to be a factor of 1.5--2 times higher than the average value
in the high Galactic latitude diffuse atomic ISM \citep{planck2013-p06b}. 
We define the ratio between the dust opacity in the
molecular and atomic media, $R$, and solve for $X_{\rm CO}$ as follows
\begin{equation}
X_{\rm CO}=\frac{\tau_{353}} { I_{\rm CO}}\frac{N_{\hi}}{2\tau_{353}} \frac{1}{R}
\label{eq:xco}
\end{equation}
where $\tau_{353} / N_{\hi} = \sigma_{\hi} =  7\times10^{-27}$\,cm$^{2}$\,H$^{-1}$
\citep{planck2013-p06b,planck2013-XVII} and $\sigma_{\rm H_{2}} = R \sigma_{\hi}$. We remove
the contribution of the atomic medium to the dust optical depth using
the $N_{\hi}$ data from the GASS survey (Sect. \ref{subsec:hidata}) and the above value of
$\sigma_{\hi}$. The GASS data only cover a fraction of the
region under study, $l=20\degr$--$36\fdg5$ at $b=0\degr$, which is nevertheless
sufficient to derive the correlation between CO emission and dust optical
depth.

\begin{figure}
\centering
\includegraphics[width=1\columnwidth]{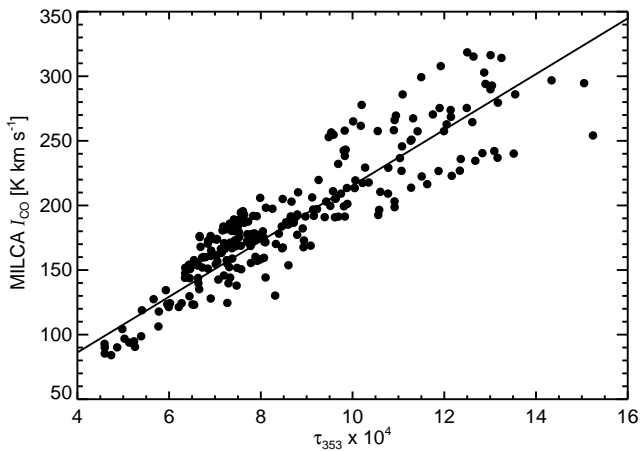}
\caption{The MILCA CO line intensity as a function of the dust
  optical depth $\tau_{353}$, corrected for the atomic gas
  contribution, along $b=0\degr$ in the $l=20\degr$--$36\fdg5$ region. The line represents
  the linear fit to the points, from which the conversion factor 
  $X_{\rm CO}=2.1\times10^{20} \times (2/R)$\,cm$^{-2}$\,(K\,\kms)$^{-1}$ is derived
(see text).}
\label{fig:tau-co}
\end{figure}

The distribution of $ I_{\rm CO}$ as a function of the \hi-corrected
$\tau_{353}$ is shown in Fig. \ref{fig:tau-co}. A linear fit to
the data passing through the origin, combined with Eq. (\ref{eq:xco}), gives $X_{\rm
  CO}=1.7\times10^{20} \times (2/R)$\,cm$^{-2}$\,(K\,\kms)$^{-1}$. The
uncertainty on this value is of 13\,\%, estimated from the scatter
of the points. In order to assess the effect of a possible
  underestimation of the true column density of the atomic gas
  (Sect. \ref{subsec:hidata}), we scale $N_{\hi}$ by a factor of 1.5 and
  repeat the analysis. We find that the uncertainties on the $N_{\hi}$
  template do not affect $X_{\rm CO}$ by more than $\sim9$\,\%.
We note that the MILCA CO data in this
region of the Galactic plane are about 25\,\% higher than the CO data from
\citet{Dame:2001} (Sect.~\ref{sec:planckdata}). Since the $X_{\rm CO}$ values in the
literature refer to the \citeauthor{Dame:2001} data, we scale our result
by 25\,\% which gives $X_{\rm CO}=2.1\times10^{20} \times
(2/R)$\,cm$^{-2}$\,(K\,\kms)$^{-1}$. If we assume $R=2$ then we obtain $X_{\rm
  CO}= X_{\rm
  CO}^{\rm ref} = 2.1\times10^{20}$\,cm$^{-2}$\,(K\,\kms)$^{-1}$, which is
consistent with the recommended value for the Galaxy given by \citet{Bolatto:2013}.

\begin{figure}
\centering
\vspace{-2.7cm}
\includegraphics[scale=0.4]{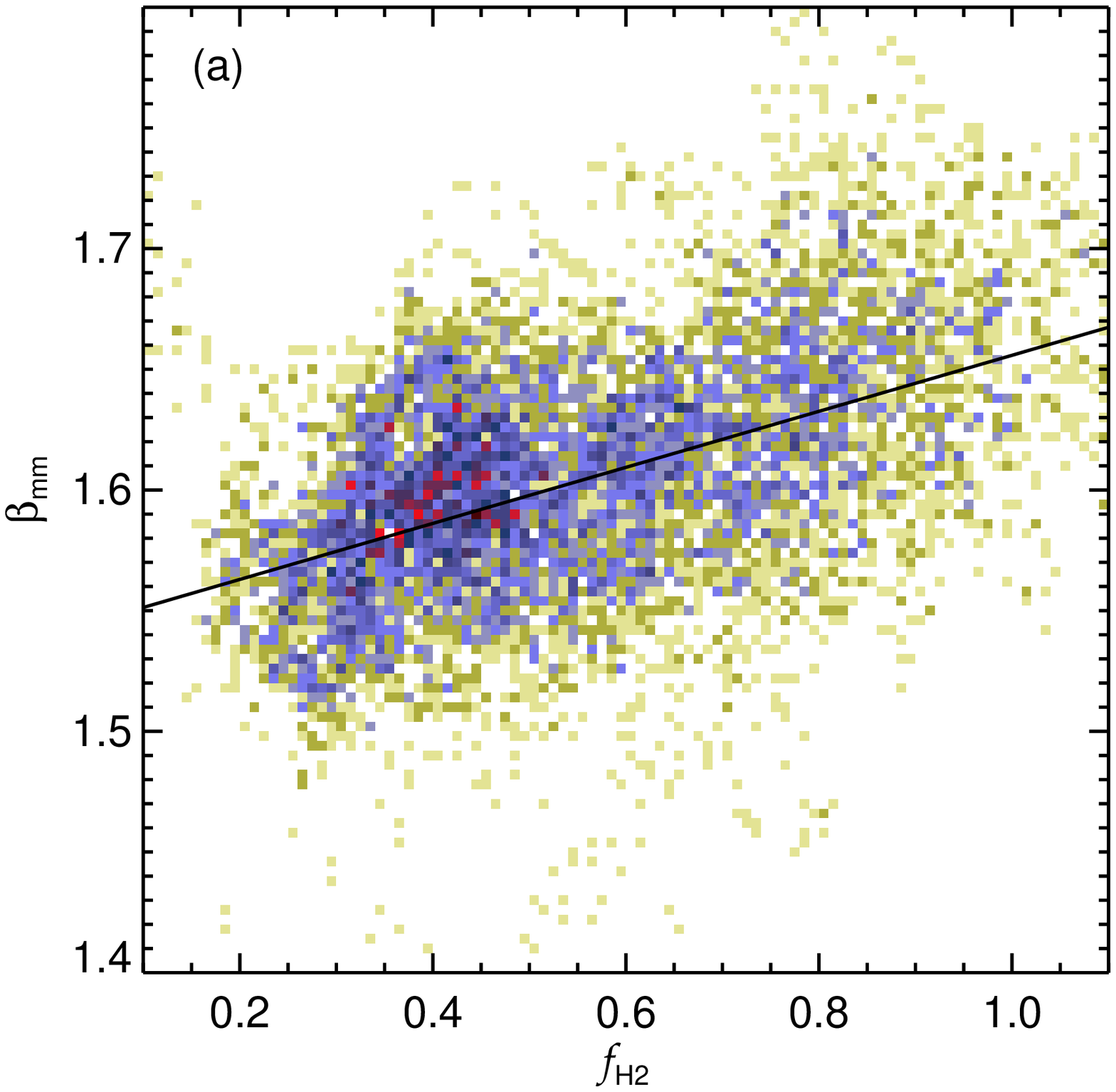}\vspace{-3cm}
\includegraphics[scale=0.4]{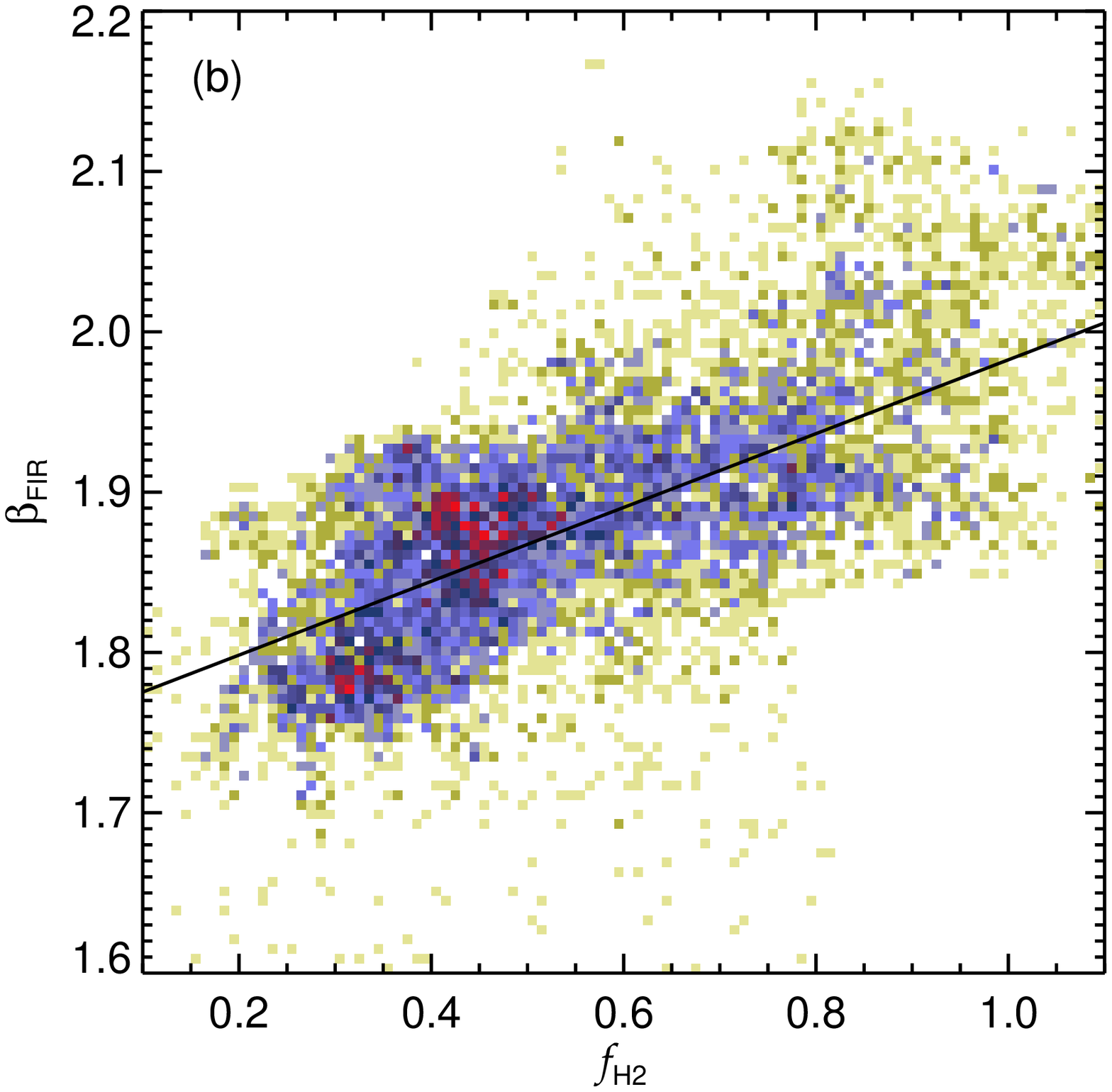}
\caption{Opacity spectral indices as a function of the fraction
  of molecular gas along the
  line of sight. (a) $\beta_{\rm mm}$ versus $f_{\rm H_{2}}$. (b)
  $\beta_{\rm FIR}$ versus $f_{\rm H_{2}}$. The points correspond to the thin Galactic
  disk, $|b| \lsim 1\degr$ or $\tau_{353} \geq 4\times 10^{-4}$. Here $f_{\rm H_{2}}$ is estimated assuming that the dust
  opacity in the molecular phase is twice as that of the atomic
  medium, $R=2$, and using $ X_{\rm CO}=2.1 \times
  10^{20}$\,cm$^{-2}$\,(K\,\kms)$^{-1}$.}
\label{fig:beta-fh2}
\end{figure}

We can write the fraction of molecular gas along the line of sight
as a function of $X_{\rm CO}$, or $R$, as
\begin{equation}
f_{\rm H_{2}}= \frac{2  I_{\rm CO} X_{\rm CO}^{\rm ref} (2/R)}{\tau_{353}/ \sigma_{\hi} + 2 I_{\rm CO}  X_{\rm CO}^{\rm ref}(2/R)(1-R) }.
\label{eq:fh2}
\end{equation}
The correlation between $\beta_{\rm
  mm}$ and $f_{\rm H_{2}}$ for $R=2$ is shown in Fig. \ref{fig:beta-fh2}(a), where $\beta_{\rm mm}$ is seen
to increase from atomic to molecular dominated regions. A linear
fit to the data gives $\beta_{\rm mm}=(1.54\pm0.01) + (0.12\pm0.01)
f_{\rm H_{2}}$, meaning that $\beta_{\rm mm}=1.54$ and 1.66 at low and high
values of $f_{\rm H_{2}}$, respectively. If we assume $R=1.5$ then $X_{\rm
    CO}=2.8\times10^{20}$\,cm$^{-2}$\,(K\,\kms)$^{-1}$, which is
  somewhat higher than the typical values for the Galaxy
\citep{Bolatto:2013}. Nevertheless, the correlation between
$\beta_{\rm mm}$ and $f_{\rm H_{2}}$, $\beta_{\rm mm}=(1.53\pm0.01) + (0.13\pm0.01)
f_{\rm H_{2}}$ is essentially unchanged. We note that we have assumed a single $T_{\rm d}$ value in the SED fit,
ignoring the fact that $T_{\rm d}$ is likely to be systematically
lower in molecular clouds than in the diffuse atomic medium
\citep{planck2011-7.0,planck2013-p06b}. However, since $T_{\rm d}$ and $\beta_{\rm mm}$ are
anti-correlated, using a lower $T_{\rm d}$ in the fit would result in
an even higher $\beta_{\rm mm}$ in molecular media. This would thus
increase the difference in $\beta_{\rm mm}$ between atomic and
molecular dominated regions. We also note that unphysical values of $f_{\rm H_{2}}$
greater than one can be reached due to data noise as well as to the
assumed values of $\sigma_{\hi}$ and $R$. In Fig. \ref{fig:beta-fh2},
where $R=2$ and $\sigma_{\hi} =  7\times10^{-27}$\,cm$^{2}$\,H$^{-1}$,
only 4\,\% of the points have $f_{\rm H_{2}} > 1$.

Recent results from \citet{Tabatabaei:2014} show an
  increase of the opacity spectral index, $\beta_{\rm FIR}$, from the outer to the inner disk
  of M33. In addition, they find this trend to be
  associated with tracers of star formation and molecular gas. Their
  analysis is based on data between 70 and 500\,$\mu$m, which they fit
  with a single and a double, cold and warm, component model. The
  results on $\beta_{\rm FIR}$ from both models are
  consistent. We find that $\beta_{\rm
  FIR}$ is also linearly correlated with $f_{\rm H_{2}}$ in the
Galactic plane, as shown in Fig. \ref{fig:beta-fh2}(b). A linear fit to the data gives $\beta_{\rm
  FIR} = (1.75\pm0.01) + (0.23\pm0.01) f_{\rm H_{2}}$, which translates into an
increase of $\beta_{\rm FIR}$ from 1.75 in atomic medium to 1.98 in
molecular medium. This result compares with that found in M33, even if the wavelength range covered in our analysis does not
allow the comprisal of two dust components in the fit. The same trend is found by
  \citet{Draine:2014} in M31, where $\beta_{\rm FIR}$ increases from $\sim 2.0$ at a distance of 15\,kpc from the centre of
  the galaxy to $\sim 2.3$ at 3\,kpc.

%________________________________________________________________
\subsection{Dust evolution}
\label{subsec:preconc}

The flattening of the dust emission towards
millimetre wavelengths in the plane of the Galaxy found here, is
accompanied by the results of \citet{planck2013-XVII}, where the same phenomenon
is detected at high Galactic latitudes. In the present work $\beta_{\rm FIR}
- \beta_{\rm mm} \sim 0.2$ for the atomic ISM, comparable to the value
measured at high Galactic latitudes. One possibility to
  explain the observed change in spectral index is an increasing contribution from carbon dust
at millimetre wavelengths. In the model of \citet{Jones:2013} the spectral
index of the carbon dust emission at 1\,mm depends on the degree of
hydrogenation and aromaticity of the grains.

It is not clear if the change of spectral index $\beta_{\rm
    mm}$  from atomic to
  molecular media is related to the observed variation of the dust specific opacity
\citep{planck2013-p06b}. Grain coagulation is a possible interpretation of the
change in dust opacity. Coagulation models indicate that dust
aggregation produces an overall increase of the dust specific opacity
in molecular clouds, without significantly changing the apparent $\beta_{\rm mm}$
\citep{Kohler:2012}. Further, there is evidence of variations in the
dust opacity within
the local atomic ISM \citep{planck2011-7.12,planck2013-XVII}, where grain
coagulation is unlikely to occur.
%, of comparable
%magnitude to what we report here. }

Further studies are needed to explain the correlation with the
  molecular material observed here, which remains phenomenological and
whose origin does not rely on a physical model. In particular,
\Planck~polarization data will be a first test of the nature of the
dust SED flattening.

%________________________________________________________________
%________________________________________________________________
\section{Conclusions}
\label{sec:conc}

We have used \Planck~HFI data to derive the power-law index of the
interstellar dust opacity
in the frequency range 100 to 353\,GHz, in a $24\degr (l) \times
8\degr (b)$ region of the Galactic plane. This is possible to achieve after the
removal of the free-free emission contribution at these frequencies,
which can be as high as 20--40\,\% of the total emission in the thin and
ionized disk of the Galaxy. Here we summarize our results:
\begin{itemize}

\item The spectral index of the dust opacity in the millimetre
  wavelength range, $\beta_{\rm mm}$, and in the Galactic plane has a median value of $1.60 \pm
  0.06$. Thus $\beta_{\rm mm}$ is smaller than that at FIR
  frequencies, $\beta_{\rm FIR}$, for which we determine a median
  value of $1.88 \pm 0.08$.

\item We find that there is no apparent trend of $\beta_{\rm mm}$ with
  temperature, as opposed to $\beta_{\rm FIR}$, for which the
  anti-correlation has been examined in several previous studies. 

\item We find that $\beta_{\rm mm}$ is, however, correlated with the derived
  dust optical depth at 353\,GHz. We interpret this correlation as an
  evolution of $\beta_{\rm mm}$ with the fraction of molecular gas along
  the line of sight, $f_{\rm H_{2}}$. Within this scenario, $\beta_{\rm mm} \sim 1.54$
  when the medium is mostly atomic, whereas it increases to about 1.66
  when the medium is predominantly molecular.

\item The results on $\beta_{\rm mm}$ are compared with predictions from two
  different physical models, TLS and emission by ferromagnetic grains, which have been suggested to explain the
  flattening of the dust emission observed at long wavelengths. We
  find that both models can in principle explain the results. The same
  applies to the standard, two dust component models, such as DL07 and
  DustEM. 

\end{itemize}

These results are important for understanding the dust emission
from FIR to millimetre wavelengths. They are key for Galactic
component separation, in particular for determining the spectral
shape of the AME at  high frequencies. Knowledge of the dust
spectrum is also critical for estimates of the free-free emission from
microwave CMB data.

\begin{acknowledgements}
We acknowledge the use of the {\tt HEALPix} \citep{Gorski:2005}
package and {\it IRAS} data. The Planck Collaboration acknowledges the
support of: ESA; CNES and CNRS/INSU-IN2P3-INP (France); ASI, CNR, and
INAF (Italy); NASA and DoE (USA); STFC and UKSA (UK); CSIC, MICINN and
JA (Spain); Tekes, AoF and CSC (Finland); DLR and MPG (Germany); CSA
(Canada); DTU Space (Denmark); SER/SSO (Switzerland); RCN (Norway);
SFI (Ireland); FCT/MCTES (Portugal); and DEISA (EU). A detailed
description of the Planck Collaboration and a list of its members can
be found at
\url{http://www.rssd.esa.int/index.php?project=PLANCK&page=Planck_Collaboration}
 The research leading to these results has received funding from the European Research Council under the European Union’s Seventh Framework Programme (FP7/2007-2013)/ERC grant agreement n$^{\circ}$ 267934.
\end{acknowledgements}

\bibliographystyle{aa}
\bibliography{Planck_bib,refs1}

\raggedright

\end{document}